\documentclass[aps,showpacs,showkeys]{revtex4} 
 
\usepackage{amsfonts}
\usepackage{amsmath}
\usepackage{amssymb}
\usepackage{color}
\usepackage{graphicx}

\usepackage[colorlinks=true,allcolors=blue]{hyperref}

\begin{document}

\title{Coherent states in the symmetric gauge for graphene\\ under a constant perpendicular magnetic field}

\author{E. D\'iaz-Bautista\footnote{ORCID: 0000-0002-2180-3895}}
\email{ediazba@ipn.mx}
\affiliation{Departamento de Formaci\'on B\'asica Disciplinaria, Unidad Profesional Interdisciplinaria de Ingenier\'ia Campus Hidalgo del Instituto Polit\'ecnico Nacional, Pachuca: Ciudad del Conocimiento y la Cultura, Carretera Pachuca-Actopan km 1+500, 42162, San Agust\'in Tlaxiaca, Hidalgo, Mexico}
\author{J. Negro\footnote{ORCID: 0000-0002-0847-6420}}
\email{jnegro@fta.uva.es}
\affiliation{Departamento de F\'{\i}sica Te\'{o}rica, At\'{o}mica y \'{O}ptica and IMUVA,
	Universidad de Valladolid, 47011 Valladolid, Spain}
\author{L. M. Nieto\footnote{ORCID: 0000-0002-2849-2647}}
\email{luismiguel.nieto.calzada@uva.es}
\affiliation{Departamento de F\'{\i}sica Te\'{o}rica, At\'{o}mica y \'{O}ptica and IMUVA,
	Universidad de Valladolid, 47011 Valladolid, Spain}

\begin{abstract}
	In this work we describe semiclassical states in graphene under a constant perpendicular magnetic field by constructing coherent states in the Barut-Girardello sense. Since we want to keep track of the angular momentum, the use of the symmetric gauge and polar coordinates seemed the most logical choice. Different classes of coherent states are obtained by means of the underlying algebra system, which consists of the direct sum of two Heisenberg-Weyl algebras. The most interesting cases are a kind of partial coherent states and the coherent states with a well-defined total angular momentum.
\end{abstract}

\maketitle

\centerline{(\today)}

\section{Introduction}\label{sec1}
The system of a charged particle in a magnetic field is, together with the harmonic oscillator, one of the most studied problems in quantum mechanics.~However,  it is still the center of a renewed interest due to its recent applications in quantum dots and other active research fields~\cite{h93,kat01,mc94,cac07,spba04}. Fock was the first  to find the solution for the physical problem of a spinless charged particle moving in the $x-y$ plane, under the simultaneous action of both, a uniform perpendicular magnetic field $\vec{B}$, and an isotropic oscillator potential $V(x,y)$. The minimal coupling time independent Schr\"{o}dinger Hamiltonian of this physical system~\cite{f28,d31} reads in the International System of Units (SI):
\begin{equation}
	\label{1}
	H=\frac{1}{2M}\left(\vec{p}- {q}\vec{A}(x,y)\right)^2+V(x,y), \quad V(x,y)=M\omega_0^2\, \frac{x^2+y^2}{2},
\end{equation}
where $M$ and $q$ are, respectively, the mass and charge of the quantum particle and, according to the so-called symmetric gauge \cite{d31,p30}, the vector potential is 
\begin{equation}
	\label{2}
	\vec{A}=\frac12\vec{B}\times\vec{r}=\frac{B_0}{2}(-y,x,0), \quad \vec{B}=\nabla\times\vec{A}=B_0\hat{k}.
\end{equation} 
Landau solved the problem \eqref{1} for $V=0$ by choosing the gauge $\vec{A}=B_0(-y,0,0)$, which nowadays is named after him, and introduced the so-called Landau levels~\cite{landau30}. His work revealed that under precise considerations, the study of a charge in a constant magnetic field reduces to solving the harmonic oscillator equation. Therefore, as Malkin and Man'ko found~\cite{mm69}, it is natural to build its coherent states as two-dimensional generalizations of Glauber's ones \cite{g63}. After these results, many research lines were developed focusing on different aspects of two-dimensional coherent states~\cite{fk70,lms89,krp96,sm03,kr05,re08,d17} and the importance of magnetic translation operators \cite{z64,b64,l83,wz94,fw99}.

On the other hand, it is well known that graphene is a material that since its discovery has exhibited interesting electronic properties which have motivated many publications, mainly due to their potential applications in the design of electronic devices. Basically, graphene consists in a sheet of carbon atoms arranged on a honeycomb lattice \cite{ngmzd04,ztsk05,cngpn09}, in which the dynamics of lower-energy electrons is described by a (2+1) dimensional massless Dirac-like equation with an effective velocity, $v_{\rm F}$, 300 times smaller than the velocity of light $c$, due to the existence of a linear dispersion relation close to the Dirac points. Thus, under these conditions, electrons in graphene behave as zero-mass Dirac particles and give rise to many relativistic phenomena, such as Klein tunneling \cite{kng06}, Hall efect \cite{ngmzd04,cngpn09,s91}, and Zitterbewegung \cite{k06,rz07,rz08}. The interaction of conducting electrons of graphene with magnetic or electric fields, as a way of controlling or confining them, has attracted growing interest. In particular, many authors have addressed the magnetic confinement of electrons in many different configurations, like square well magnetic barriers \cite{dmdae07,dndm09}, radial magnetic fields \cite{gmr09}, magnetic fields corresponding to solvable potentials \cite{knn09,mf14}, smooth inhomogeneous magnetic fields \cite{rvp11,lara,dp16,cdmp16,ema17,rkb12,dnvhp17}, etc. In this context, following Malkin and Man'ko's ideas \cite{mm69}, one can try to build the coherent states for such a kind of systems considering, in principle, homogeneous perpendicular magnetic fields. A first attempt in that direction was given in \cite{df17}, where coherent states were constructed assuming the Landau gauge $\vec{A}=B_0x\,\hat{\j}$, and working with the time-independent Dirac-Weyl (DW) equation near to one of the Dirac points, namely $K$,
\begin{equation}\label{3}
	H_{\rm DW}\Psi(x,y)=v_{\rm F}\,\vec{\sigma}\cdot\left(\vec{p}+e\vec{A}\right)\Psi(x,y)=E\Psi(x,y),
\end{equation}
being $\vec{\sigma}=(\sigma_x,\sigma_y,\sigma_z)$ the Pauli matrices and $q=-e$ the charge of the electron ($e>0$). In this situation, the coherent states are described by wave functions that correspond to a system that has a translational symmetry along the $y$ direction. 

In the present work we want to build coherent states of graphene under a constant magnetic field in the sense of Barut-Girardello \cite{bg71}, but we will study their rotational invariance by means of the symmetric gauge \eqref{2}.
In Section~\ref{sec2}, the Dirac-Weyl equation \eqref{3} in the symmetric gauge and its associated algebraic structure are discussed, in particular its energy spectrum and eigenfunctions. In Section~\ref{sec3}, families of partial and two-dimensional coherent states in graphene are obtained as eigenstates of two independent generalized annihilation operators, $\mathbb{A}^-$ and $\mathbb{B}^-$. The corresponding probability and current densities, as well as the mean energy are also evaluated. In Section~\ref{sec4}, coherent states with a fixed total angular momentum are built as eigenstates of the operator $\mathbb{K}^-=\mathbb{A}^-\mathbb{B}^-$. Our final conclusions are presented in Section~\ref{sec5}.

\section{Dirac-Weyl Hamiltonian}\label{sec2}

Using the symmetric gauge  given in (\ref{2}), the stationary DW equation~(\ref{3}) is rewritten as
\begin{equation}\label{4}
	H_{\rm DW}\Psi(x,y)=v_{\rm F}\left(\sigma_x\left(p_x-\frac{eB_0}{2}y\right)+\sigma_y\left(p_y+\frac{eB_0}{2}x\right)\right)\Psi(x,y)=E\Psi(x,y).
\end{equation}
If we introduce the magnetic length parameter ($\ell_{\rm B}$) and the so-called cyclotron frequency in this context ($\omega$) as
\begin{equation}\label{4ymedio}
	\ell_{\rm B}=\sqrt{\frac{\hbar}{eB_0}}, \qquad \omega= \frac{\sqrt{2}\,v_{\rm F}}{\ell_{\rm B}}, 
\end{equation}
Eq.~\eqref{4} can be expressed in the form
\begin{equation}\label{5}
	H_{\rm DW}\Psi(x,y)= \hbar\,\omega \left[\begin{array}{cc}
		0 & -iA^- \\
		iA^+ & 0
	\end{array}\right]\Psi(x,y)=E\Psi(x,y),
\end{equation}
where the pseudo-spinor eigenfunctions are chosen as 
\begin{equation}\label{seis}
	\Psi(x,y)=\left(\begin{array}{c}
		\psi_1(x,y) \\
		i\,\psi_2(x,y)
	\end{array}\right),
\end{equation}
and the mutually adjoint operators $A^\pm$, satisfying the commutation relation that corresponds to the Heisenberg-Weyl algebra of the harmonic oscillator
\begin{equation}\label{7}
	[A^-,A^+]=1,
\end{equation}
are defined by
\begin{equation}\label{6}
A^\pm=\mp\frac{i\, \ell_{\rm B}}{\sqrt{2}\,\hbar}\left(\left(p_x-\frac{\hbar}{2\ell_{\rm B}^2}y\right)\pm i\left(p_y+\frac{\hbar}{2\ell_{\rm B}^2} x\right)\right).
\end{equation}
Then, the eigenvalue equation (\ref{5}) gives rise to two coupled equations:
\begin{eqnarray}
	A^-\psi_2(x,y)&=&\epsilon\psi_1(x,y), \qquad 
	A^+\psi_1(x,y)=\epsilon\psi_2(x,y), \qquad  \epsilon\equiv E/ (\hbar\,\omega). 
\end{eqnarray}
After decoupling the expressions above, we obtain the following dimensionless equations for each pseudo-spinor component
\begin{eqnarray}
	\mathcal{H}_1\psi_1(x,y)=A^-A^+\psi_1(x,y)=\mathcal{E}_1\, \psi_1(x,y), \label{8a} \qquad
	\mathcal{H}_2\psi_2(x,y)=A^+A^-\psi_2(x,y)=\mathcal{E}_2\, \psi_2(x,y), 
\end{eqnarray}
where $\mathcal{H}_1$, $\mathcal{H}_2$ are effective Schr\"{o}dinger-like Hamiltonians and the effective energy is
\begin{equation}
	\label{xyz}
	\mathcal{E}_1=\mathcal{E}_2=\epsilon^2=\left(\frac{E}{\hbar\,\omega}\right)^2.
\end{equation}
Due to (\ref{7}), expressions \eqref{8a}--\eqref{xyz} are in fact the equations of two displaced harmonic oscillators, $\mathcal{H}_1=\mathcal{H}_2+1$, with energies given by
\begin{equation}\label{13}
	\mathcal{E}_{1,n-1}=\mathcal{E}_{2,n}=n, \quad n\geq 1, \quad \mathcal{E}_{2,0}\equiv0,
\end{equation} 
so that spectrum of the DW equation (\ref{5}) is
\begin{equation}\label{14}
	E_n=\pm\hbar\,\omega\, \sqrt{n}, \quad n=0,1,2,\dots,
\end{equation}
with the positive (negative) sign corresponding to the conduction (valence) band, and $\omega$ the cyclotron frequency given in \eqref{4ymedio}.

\subsection{Algebraic treatment}\label{sec2.1}	
Next, we want to construct the eigenfunctions in an algebraic way by computing the symmetries and other relevant operators. Since the problem has a geometrical rotational symmetry around the $z$-axis, it is convenient to express the Hamiltonians $\mathcal{H}_j$, $j=1,2$, together with other operators in polar coordinates $(r,\theta)$. Thus, 
\begin{equation}\label{10}
\mathcal{H}_j= \frac{\ell_{\rm B}^{2}}{2} \left(-\left(\partial_r^2+\frac{1}{r}\partial_r+\frac{1}{r^2}\partial_\theta^2\right)-\frac{i}{\ell_{\rm B}^{2}}\partial_\theta+\frac{1}{4\ell_{\rm B}^{4}}r^2\right)+\frac{(-1)^{j-1}}{2}.
\end{equation}
By introducing the dimensionless variable $\xi$ defined as
\begin{equation}\label{11}
\xi=\frac{r}{\sqrt{2}\ell_{\rm B}},
\end{equation}
the corresponding eigenvalue equations take the form
\begin{equation}\label{12}
	\mathcal{H}_j\psi_j(\xi,\theta)=\frac{1}{4}\left(-\left(\partial_\xi^2+\frac{1}{\xi}\partial_\xi+\frac{1}{\xi^2}\partial_\theta^2\right)-2i\partial_\theta+\xi^2+2(-1)^{j-1}\right)\psi_j(\xi,\theta)=\mathcal{E}_j\psi_j(\xi,\theta).
\end{equation}
This set of differential equations reminds the well known Fock-Darwin system \cite{f28,d31,dknn17}. Both Hamiltonians $\mathcal{H}_j$ can also be factorized in terms of two new differential operators $B^\pm$ that are obtained following the factorization procedure given in \cite{df96,kka12}:
\begin{equation}\label{15}
B^\pm=\frac{\exp(\mp i\theta)}{2}\left(\mp\partial_\xi+\frac{i\partial_\theta}{\xi}+\xi\right)=\mp\frac{i\ell_{\rm B}}{\sqrt{2}\hbar} \left(\left(p_x+\frac{\hbar}{2\ell_{\rm B}^2} y\right)\mp i\left(p_y-\frac{\hbar}{2\ell_{\rm B}^2} x\right)\right).
\end{equation}
Then, it is easily checked that
\begin{equation}\label{17}
\mathcal{H}_1=B^-B^++L_z, \qquad \mathcal{H}_2=B^+B^-+L_z,
\end{equation}
where $L_z=\left(xp_y-yp_x\right)/\hbar=-i\partial_{\theta}$
denotes the $z$-component of the angular momentum operator in cartesian coordinates.

The two operators $B^{\pm}$ constitute a second set of boson operators that commute with the previous set, $A^{\pm}$ given in (\ref{6}), which in polar coordinates have the following expressions
	\begin{equation}\label{35}
A^\pm=\frac{\exp(\pm i\theta)}{2}\left(\mp\partial_\xi-\frac{i\partial_\theta}{\xi}+\xi\right).
\end{equation}
Therefore,
\begin{equation}\label{16}
[B^-,B^+]=1, \qquad [A^\pm,B^\pm]=0, \qquad [A^\pm,B^\mp]=0.
\end{equation}
From the factorizations (\ref{8a}) in terms of $A^{\pm}$ and (\ref{17}) in terms of $B^{\pm}$, it follows that $L_z$ can be expressed as
\begin{equation}\label{19}
	L_z=A^+A^--B^+B^-
\end{equation}
and satisfies the commutation relations
\begin{equation}\label{20}
	[L_z,A^\pm]=\pm A^\pm, \qquad [L_z,B^\pm]=\mp B^\pm.
\end{equation}
This implies that $A^{+}$ increases and $A^{-}$ decreases the eigenvalues of each $\mathcal{H}_{j}$ in one unit so that they act as ladder operators. On the other hand, the two operators $B^\pm$ commute with both $\mathcal{H}_1$ and $\mathcal{H}_2$, and constitute a pair of symmetries. The operators $B^\pm$ are related to the so-called {\it magnetic translation operators} \cite{z64,b64,l83}, which  generate the translation of the center of the classical circular orbits. This fact will be discussed in the following section for the first family of partial coherent states.
In addition, it is easily checked that the operators $A^+$ and $A^-$, acting on an eigenstate of $L_z$, 
respectively increases or decreases its eigenvalue in one unity; the operators $B^\pm$ have the opposite effect.

\subsection{Eigenstates}\label{sec2.2}
\begin{figure}[htb]
	\centering
	\includegraphics[width=.34\linewidth]{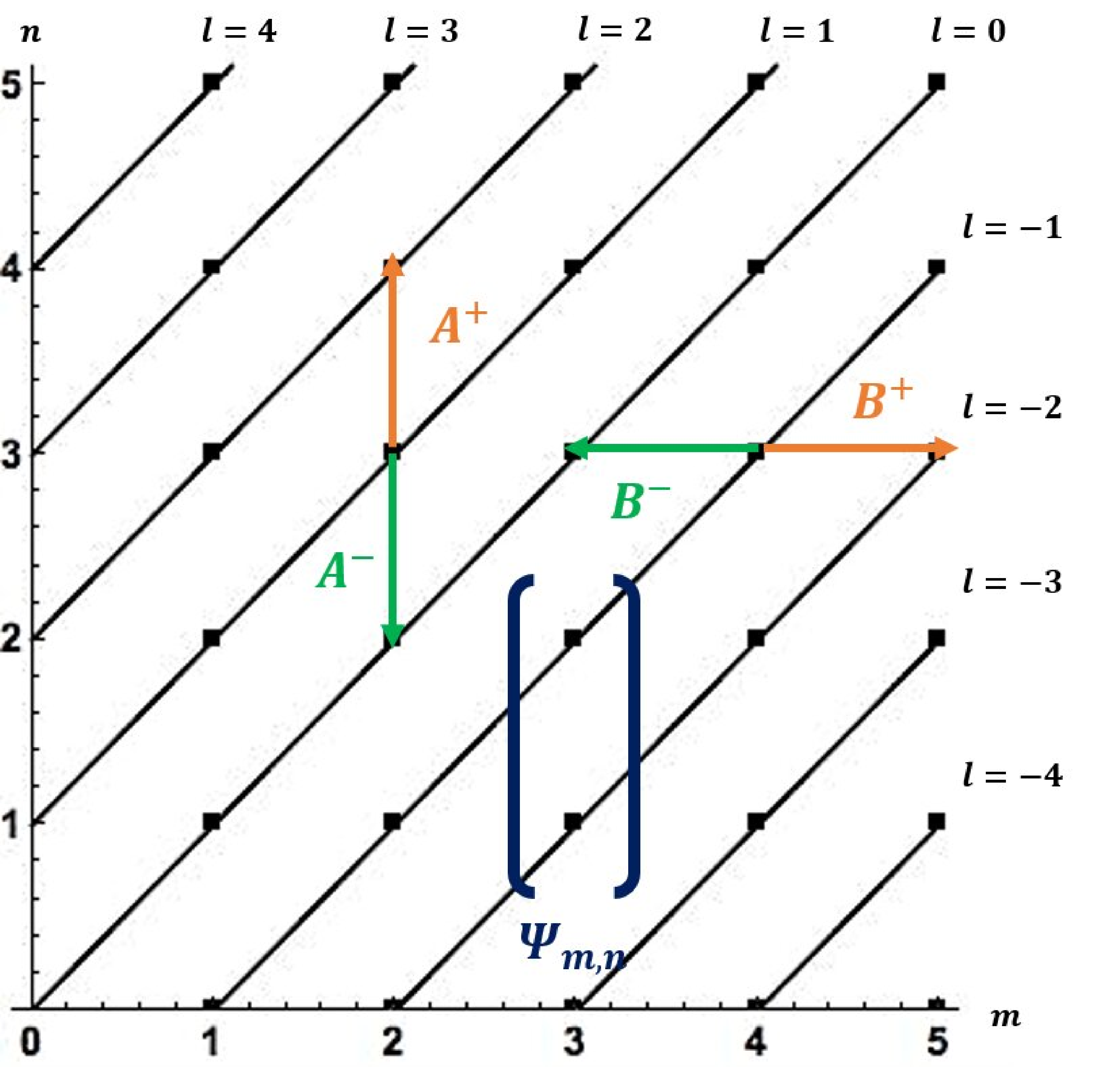}
	\caption{\label{fig:diagram}Diagram showing the space of scalar states $\psi_{m,n}$ \eqref{26a} and their connections through the action of the operators \eqref{24a}--\eqref{24b}. Each point $(m,n)$ identifies just one state $\psi_{m,n}$. The tilted lines connect states with the same eigenvalue $l=n-m$.}
\end{figure}

Now, we consider the corresponding number and angular momentum operators
\begin{equation}\label{22}
N\equiv A^+A^-, \qquad M\equiv B^+B^-, \qquad L_z\equiv N-M,
\end{equation}
which commute among themselves. Therefore, the eigenstates $\psi_j$ of the Hamiltonians $\mathcal{H}_j$ can be labeled by means of two positive integer numbers $m,n\in\mathbb{Z}^+\cup\{0\}$ that correspond to the eigenvalues of the number operators $M$ and $N$, respectively. Then, for $\psi_j\equiv\psi_{m,n}$ we have
\begin{equation}\label{23}
  N\,\psi_{m,n}=n\,\psi_{m,n}, \qquad M\,\psi_{m,n}=m\,\psi_{m,n}, \qquad L_z\,\psi_{m,n}=(n-m)\,\psi_{m,n},
\end{equation}
where the last equation implies that $\psi_{m,n}$ are also eigenstates of the operator $L_z$ with eigenvalue $l\equiv n-m$. Hence, the eigenvalue equations (\ref{8a}) of the effective Hamiltonians $\mathcal{H}_{j}$ for these number states are
\begin{equation}\label{26a}
\mathcal{H}_{1}\psi_{m,n-1}=n\psi_{m,n-1}, \qquad \mathcal{H}_{2}\psi_{m,n}=n\psi_{m,n}.
\end{equation} 
We can say that label $n$ fixes the energy and label $m$ the (infinite) degeneracy. Moreover, the action of  operators $A^\pm$ and $B^\pm$ on the states $\psi_{m,n}$ is (see Figure~\ref{fig:diagram}):
	\begin{eqnarray}
	&A^-\psi_{m,n}=\sqrt{n}\,\psi_{m,n-1}, \quad& A^+\psi_{m,n}=\sqrt{n+1}\,\psi_{m,n+1}, \label{24a}\\
	&\;B^-\psi_{m,n}=\sqrt{m}\,\psi_{m-1,n}, \quad& B^+\psi_{m,n}=\sqrt{m+1}\,\psi_{m+1,n}. \label{24b}
	\end{eqnarray}

Taking into account (\ref{24a}), \eqref{5} and  (\ref{seis}), we can identify the pseudo-spinor eigenfunctions: the fundamental states $\Psi_{m,0}(x,y)$ of the DW equation have the form
\begin{equation}\label{32}
\Psi_{m,0}(x,y)=\left(\begin{array}{c}
0 \\ i\,\psi_{m,0}(x,y)
\end{array}\right), \quad E_0=0,
\end{equation}
while the excited states, with $n\geq1$, turn out to be
\begin{equation}\label{33}
\Psi_{m,n}(x,y)=\frac{1}{\sqrt{2}}\left(\begin{array}{c}
\psi_{m,n-1}(x,y)\\
i\,\psi_{m,n}(x,y)
\end{array}\right), \quad E_n=\pm \hbar\,\omega\, \sqrt{n}, \ 
\end{equation}
with $m=0,1,2,\dots$ From now on we will analyze only the states with $E\geq 0$. Recently, the eigenvalues and  eigenfunctions for the zero modes of the DW equation for other magnetic fields were obtained in \cite{s18,kns18} by applying the Aharonov-Casher theorem \cite{ac79}.
The states $\psi_{m,n}$ can be built from the successive action of the creation operators $A^+$ and $B^+$ on the fundamental state $\psi_{0,0}$:
\begin{equation}\label{25}
\psi_{m,n}=\frac{(B^+)^m(A^+)^{n}}{\sqrt{m!\,n!}}\, \psi_{0,0}, \quad m,\,n=0,1,2,\dots,
\end{equation}
where the ground state $\psi_{0,0}$ is determined by the conditions
\begin{equation}\label{26}
A^-\psi_{0,0}=B^-\psi_{0,0}=0.
\end{equation}
Using the polar coordinate expressions of $A^{-}$, $B^{-}$ in (\ref{15}) and (\ref{35}), the wave function of this state is found to be
\begin{equation}
\psi_{0,0}(\xi,\theta)=K_0 \ e^{-\xi^2/2},
\end{equation}
where $K_0$ is a normalization constant. To obtain the wave functions of the excited states $\psi_{m,n}(\xi,\theta)$ one can use the fact that they can be expressed as separated functions \cite{dknn17}:
\begin{equation}\label{27}
\psi_{m,n}(\xi,\theta)=  R_{m,n}(\xi)\, \Theta_l(\theta), \quad l=n-m,
\end{equation}
where $\Theta_l(\theta)$ is an eigenfuction of $L_z=-i\partial_{\theta}$, {\it i.e.},
\begin{equation}\label{28}
\Theta_l(\theta)=\exp(il\theta), \quad L_z\Theta_l(\theta)=l\, \Theta_l(\theta), \quad l=0,\pm1,\pm2,\dots,
\end{equation}
and the radial function
$R_{m,n}(\xi)$ can be written as
\begin{equation}\label{29}
R_{m,n}(\xi)=K_{m,n}\,\xi^{\vert n-m\vert}\,  \ e^{-\xi^2/2}\, f_{mn}(\xi),
\end{equation}
where $K_{m,n}$ are normalization constants and $f_{mn}(\xi)$ are functions to be determined. After the change $t=\xi^2$ and by substituting into (\ref{26a}) and (\ref{12}), we obtain the following differential equations
	\begin{eqnarray}
	&&t\, \frac{\mathrm{d}^2f_{mn}(t)}{\mathrm{d}t^2}+(1+n-m-t)\frac{\mathrm{d}f_{mn}(t)}{\mathrm{d}t}+mf_{mn}(t)=0, \quad n>m, \label{30a} \\
	&&t\, \frac{\mathrm{d}^2f_{mn}(t)}{\mathrm{d}t^2}+(1+m-n-t)\frac{\mathrm{d}f_{mn}(t)}{\mathrm{d}t}+nf_{mn}(t)=0, \quad m>n, \label{30b}
	\end{eqnarray}
with $f_{mn}(\xi)\equiv f_{mn}(t)$, whose solutions can be expressed in terms of associated Laguerre polynomials $L_k^{\alpha}(t)$.
Hence, after simple calculations, the normalized eigenfunctions of the Hamiltonian $\mathcal{H}_j$ are found to be
\begin{equation}\label{31}
   \psi_{m,n}(\xi,\theta)= \frac1{\ell_{\rm B}} (-1)^{\min(m,n)} \sqrt{\frac{1}{2\pi}\frac{\min(m,n)!}{\max(m,n)!}}
   \ \xi^{\vert n-m\vert}\  e^{-\frac{\xi^2}{2}+i(n-m)\theta}\
L_{\min(m,n)}^{\vert n-m\vert} (\xi^2 ), \quad n,m=0,1,2,\dots
\end{equation}
Observe that in this equation the only dependence on the physical constants is in the factor $1/\ell_{\rm B}$, and therefore the remaining term is a result valid  for any arbitrary constant magnetic field. 
These kind of solutions were obtained initially in \cite{f28}.
Notice that the set of eigenstates $\psi_{m,n}$, represented in the first quadrant of the plane with coordinates $(m,n)$ in Figure~\ref{fig:diagram}, is divided in two sectors, according to whether $l> 0$ (upper sector) or $l\leq0$ (lower sector). The states with $l=0$ are located in the bisector of this first quadrant. In this sense, although the pseudo-spinor eigenstates $\Psi_{m,n}(x,y)$ are composed of the two scalar states $\psi_{m,n}(x,y)$ and $\psi_{m,n-1}(x,y)$ with different value of $l$, both of them can belong to the same sector. Therefore, we can denote as $\Psi_{m,n}^{+}(x,y)$ the pseudo-spinor states whose two scalar components have positive $z$-component of the angular momentum $(l>0)$, and as $\Psi_{m,n}^{-}(x,y)$ those whose two scalar components have negative values $(l\leq0)$, {\it i.e.},
\begin{eqnarray}
\label{34aaa}
\Psi_{m,n}^{+}(x,y)&=&\frac{1}{\sqrt{2}}\left(\begin{array}{c}
\psi_{m,n-1}^{+}(x,y)\\
i\psi_{m,n}^{+}(x,y)
\end{array}\right), \quad n> m, 
\\ [1ex]
\Psi_{m,n}^{-}(x,y)&=&\frac{1}{\sqrt{2^{(1-\delta_{0n})}}}\left(\begin{array}{c}
(1-\delta_{0n})\psi_{m,n-1}^{-}(x,y)\\
i\psi_{m,n}^{-}(x,y)
\end{array}\right), \quad n\leq m,
\label{34bbb}
\end{eqnarray}
where $\delta_{mn}$ is the Kronecker delta and $\psi^+_{m,n}(x,y)$  ($\psi^-_{m,n}(x,y)$) identifies the states that belong to the upper (lower) sector in Figure \ref{fig:diagram}.

In addition, by defining the total angular momentum operator in the $z$-direction as $\mathbb{J}_z=L_z\otimes\mathbb{I}+\sigma_z/2$, we have that
\begin{eqnarray}\label{36}
 \mathbb{J}_z\,\Psi_{m,n}^{+}(x,y)=j\,\Psi_{m,n}^{+}(x,y), \qquad \mathbb{J}_z\,\Psi_{m,n}^{-}(x,y)=j\,\Psi_{m,n}^{-}(x,y), 
\end{eqnarray}
{\it i.e.}, the states $\Psi_{m,n}$ are also eigenstates of $\mathbb{J}_z$ with eigenvalue $j\equiv l-1/2$. More precisely, the states $\Psi_{m,n}^+$ have $j\geq 1/2$ and the states  $\Psi_{m,n}^-$ have  $j\leq-1/2$.

\subsubsection{Probability and current densities}\label{sec2.2.1}
\begin{figure}[htb]
	\centering
	\begin{minipage}[b]{0.34\linewidth}
		\includegraphics[width=\textwidth]{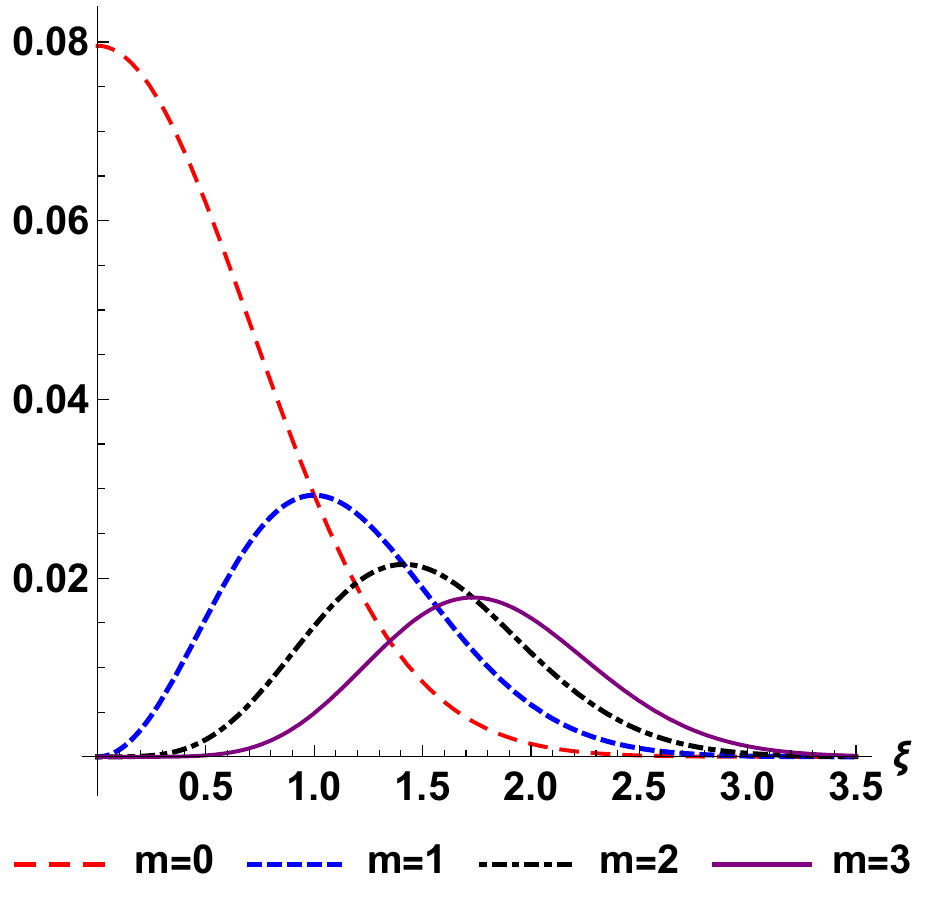}
	\end{minipage}
	\caption{\label{fig:rho_NM} Dimensionless probability density $\ell_{\rm B}^{2}\, \tilde{\rho}_{m,0}(\xi)$ in \eqref{37} for some pseudo-spinor ground states $\Psi_{m,0}$ given by \eqref{32}, with $E_0=0$ and total angular momentum in $z$-direction $j=-(m+1/2)$.}
\end{figure}

To describe the physical properties of the states  $\Psi_{m,n}$, we construct their probability and current densities in terms of the polar coordinates $(\xi,\theta)$.
The radial probability density $\tilde{\rho}_{m,n}(\xi)$ for $m,n=0,1,2,\dots$, is given by
\begin{eqnarray}
 \nonumber\tilde{\rho}_{m,n}(\xi)&=&\Psi_{m,n}^\dagger \Psi_{m,n}=\frac{\vert\psi_{m,n} \vert^2+(1-\delta_{0n})\vert\psi_{m,n-1} \vert^2}{2^{(1-\delta_{0n})}}
\\
  &=&\frac1{\ell_{\rm B}^{2}} \frac{1}{2^{(2-\delta_{0n})}\pi} \exp\left(-\xi^2\right)\left\{\frac{\min(m,n)!}{\max(m,n)!}\,\xi^{2\vert n-m\vert}\left( L_{\min(m,n)}^{\vert n-m\vert} (\xi^2 )\right)^2\right.  
\nonumber
\\
 & &\left. \qquad\qquad\qquad\qquad\qquad\quad +(1-\delta_{0n})\frac{\min(m,n-1)!}{\max(m,n-1)!}\,\xi^{2\vert n-m-1\vert}\left(L_{\min(m,n-1)}^{\vert n-m-1\vert} (\xi^2 )\right)^2\right\}.
\label{37}
\end{eqnarray}
The scalar radial probability density  corresponding to the scalar component $\psi_{m,n}$ will be denoted by $\rho_{m,n}(\xi)\equiv\vert\psi_{m,n}\vert^2$. In Figure~\ref{fig:rho_NM} plots of the radial probability density $\tilde{\rho}_{m,0}(\xi)$ for the first pseudo-spinor ground states $\Psi_{m,0}$ are shown.

\begin{figure}[htb]
	\centering
	\begin{minipage}[b]{0.3\textwidth}
		\hspace{0.1cm}
		\includegraphics[width=\textwidth]{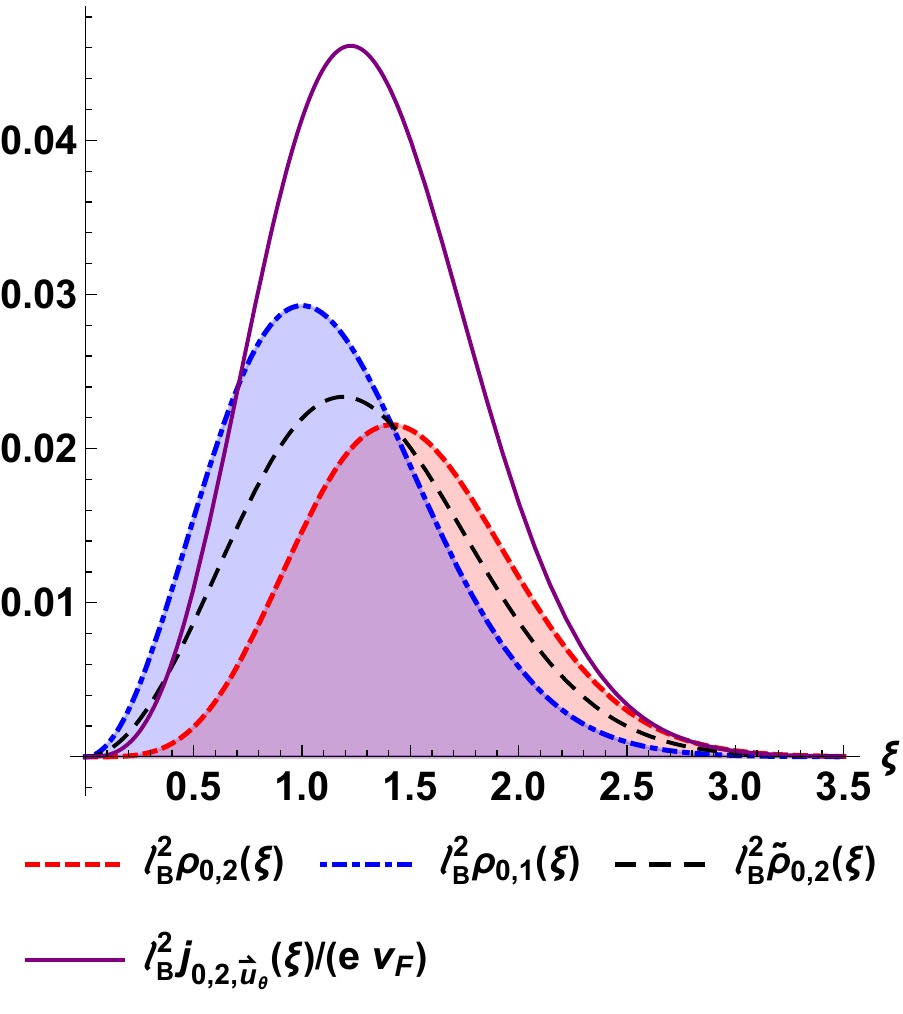}\\
		\centering{\footnotesize (a) $\Psi_{0,2}$, $j=3/2$.}
		\label{fig:rho_j_20}
	\end{minipage}
\hspace{3cm}
	~ 
	\begin{minipage}[b]{0.3\textwidth}
		\includegraphics[width=\textwidth]{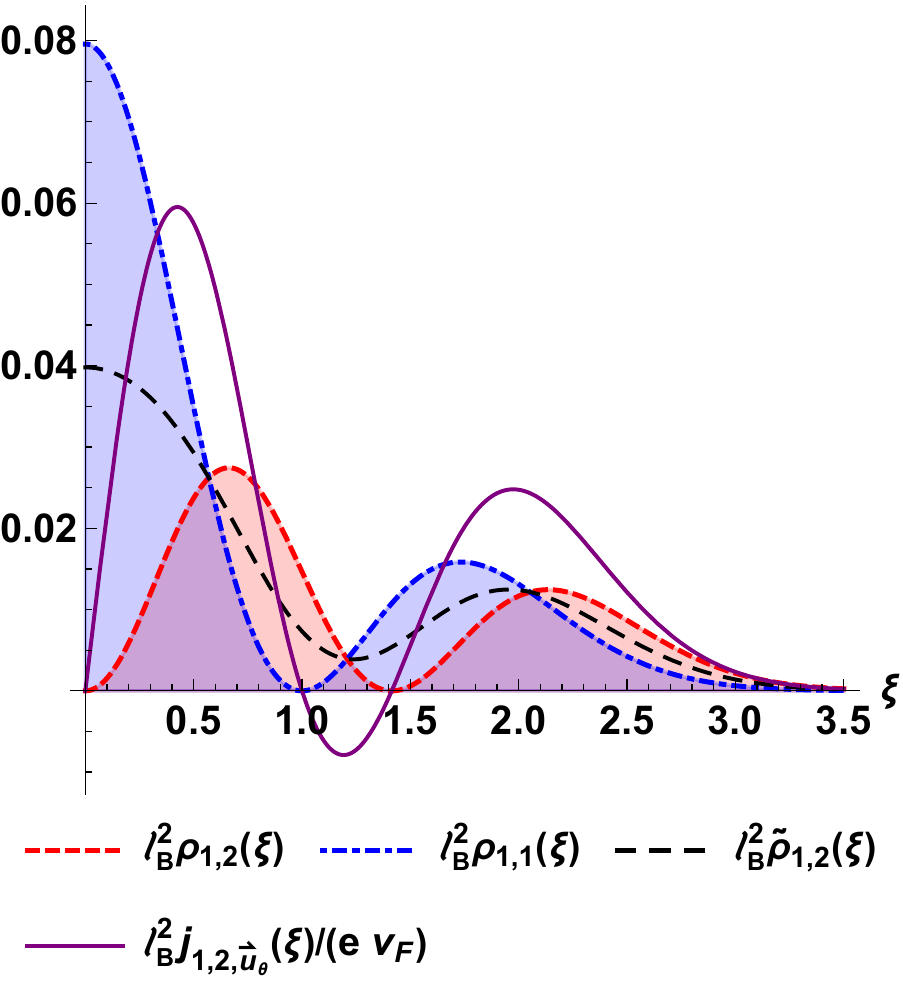}\\
		\centering{\footnotesize (b) $\Psi_{1,2}$, $j=1/2$.}
		\label{fig:rho_j_21}
	\end{minipage}
	\\ [3ex]
	~ 
	\begin{minipage}[b]{0.34\textwidth}
		\includegraphics[width=\textwidth]{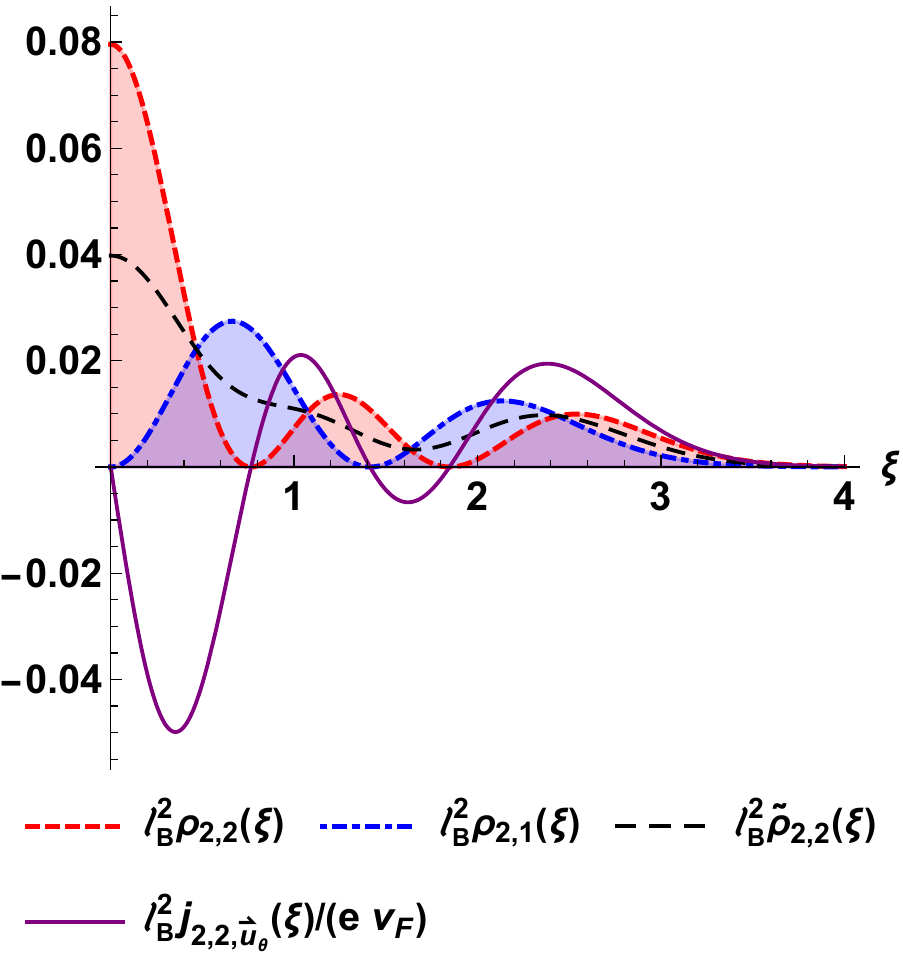}\\
		\centering{\footnotesize (c) $\Psi_{2,2}$, $j=-1/2$.}
		\label{fig:rho_j_22}
	\end{minipage}
\hspace{3cm}
	~ 
	\begin{minipage}[b]{0.34\textwidth}
		\includegraphics[width=\textwidth]{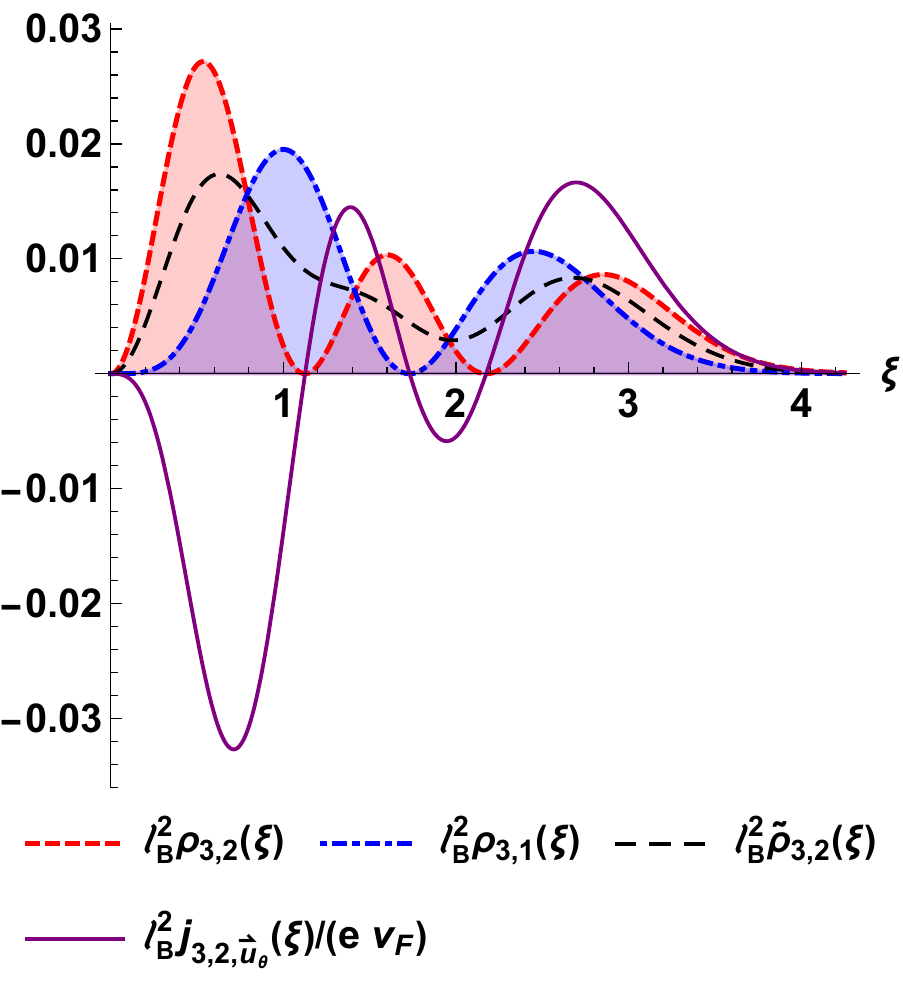}\\
		\centering{\footnotesize (d) $\Psi_{3,2}$, $j=-3/2$.}
		\label{fig:rho_j_23}
	\end{minipage}
	\caption{\label{fig:rho_j_2M}Comparison between the dimensionless probability densities $\ell_{\rm B}^{2}\, \tilde{\rho}_{m,2}(\xi)$ in \eqref{37} (black dashed lines) and current densities $\ell_{\rm B}^{2}\, j_{m,2,\vec{u}_\theta}(\xi) /(e\,v_{\rm F})$ in \eqref{39b} (purple solid lines) for the pseudo-spinor states $\Psi_{m,2}$ with energy $E=\sqrt{2}\hbar v_{\rm F}$. The scalar probability densities $\rho_{m,n}(\xi)$ associated to each pseudo-spinor component are shown in red and blue dashed lines. The corresponding value of the total angular momentum $j$ in $z$-direction is indicated in each plot.}
\end{figure}

The stationary states of the DW equation may have non-vanishing current density $j_{\vec{u}}$ in the direction of the unit vector $\vec{u}$. The proper definition of this current density for the state $\Psi_{m,n}$ is
\begin{equation}\label{38}
j_{m,n,\vec{u}}=ev_{\rm F}\Psi_{m,n}^\dagger\,(\vec{\sigma}\cdot\vec{u})\,\Psi_{m,n}.
\end{equation}
In particular, considering the directions along the polar vectors $\vec{u}_\xi$ and $\vec{u}_\theta$, we have for $m\geq0$, $n\geq1$:
\begin{eqnarray}
   j_{m,n,\vec{u}_\xi}(\xi) \!&\!=\!&\! ev_{\rm F}\Psi_{m,n}^\dagger\,\left(\vec{\sigma}\cdot\vec{u}_\xi\right)\,\Psi_{m,n}= ev_{\rm F}\Psi_{m,n}^\dagger\,\left[\begin{array}{c c}
0 & e^{-i\theta} \\
e^{i\theta} & 0
\end{array}\right]\Psi_{m,n}=0, 
\label{39a} 
\\
   j_{m,n,\vec{u}_\theta}(\xi)\!&\!=\!&\! ev_{\rm F}\Psi_{m,n}^\dagger\,(\vec{\sigma}\cdot\vec{u}_\theta)\,\Psi_{m,n}= ev_{\rm F}\Psi_{m,n}^\dagger\,\left[\begin{array}{c c}
0 & -ie^{-i\theta} \\
ie^{i\theta} & 0
\end{array}\right]\Psi_{m,n} \nonumber
\\
  \!&\!=\!&\! \frac1{\ell_{\rm B}^{2}} \textrm{sgn}(n-m)\,  \frac{ev_{\rm F}}{2\pi} \, \frac{\sqrt{\min(m,n-1)!\min(m,n)!}}{\sqrt{\max(m,n-1)!\max(m,n)!}} 
\, \xi^{\vert2n-2m-1\vert}  e^{-\xi^2}\, L_{\min(m,n)}^{\vert n-m\vert}(\xi^2) \, L_{\min(m,n-1)}^{\vert n-m-1\vert }(\xi^2). \label{39b}
\end{eqnarray}
These expressions indicate that there is no probability flux in the radial direction $\vec{u}_\xi$, while the probability density in the angular direction $\vec{u}_\theta$ is symmetric with respect to rotations around the $z$-axis. Both current densities are null for the set of fundamental states $\Psi_{m,0}$. In Figure~\ref{fig:rho_j_2M} the behavior of both probability and current densities corresponding to some states $\Psi_{m,n}$ are plotted and compared. As we can see, the probability density of the pseudo-spinor states $\Psi_{m,n}$ remains between the probability densities of their corresponding scalar components $\psi_{m,n}$. Also, as $m$ increases, the sign of the current density $j_{m,n,\vec{u}_\theta}(\xi)$ changes in the points in which the scalar densities $\rho_{m,n}(\xi)$ show a minimum value.

\section{Partial coherent states}\label{sec3}
The DW problem in graphene we are dealing with belongs to a kind of pseudo-spinor-like systems in which the solutions are expressed as wave functions of two components, as occurs with supersymmetric harmonic oscillator \cite{w81}. To apply the coherent states formalism to such a system, a supersymmetric annihilation operator must be defined in a general form. Unfortunately, it is known that it lacks uniqueness \cite{az86,bh93,kz13,df17}: there is a certain freedom to construct the coherent states associated with a specific form of the supersymmetric annihilation operator. In this sense, there are also different ways to define creation and annihilation operators for the DW pseudo-spinors starting from the scalar creation and annihilation operators $A^\pm$, $B^\pm$. For instance, let us consider the following definition of operators depending on arbitrary parameters $\delta,\,\eta\in[0,2\pi]$:
	\begin{equation}
	\mathbb{A}^-\! =\!\left[\begin{array}{c c}
	\cos\delta \frac{\sqrt{N+2}}{\sqrt{N+1}}A^- & \sin\delta \frac{1}{\sqrt{N+1}}(A^-)^2 \\ [1ex]
	-\sin\delta \sqrt{N+1} & \cos\delta A^-
	\end{array}\right]\!\!,\,   \mathbb{A}^+ \!=\! (\mathbb{A}^-)^\dagger, \label{41a}
	\ 
	\mathbb{B}^- \!=\!\left[\begin{array}{c c}
	\cos\eta\, B^- & \sin\eta\,\frac{B^-}{\sqrt{N+1}}A^- \\ [1ex]
	-\sin\eta\, A^+\frac{B^-}{\sqrt{N+1}} & \cos\eta\, B^-
	\end{array}\right]\!\!,\,  \mathbb{B}^+ \!=\!(\mathbb{B}^-)^\dagger,  
	\end{equation}
Their action on the eigenstates, as long as $n\neq0$, is quite reasonable:
\begin{equation}\label{45a}
 \mathbb{A}^{-}\Psi_{m,n+1}=e^{i\delta}\, \sqrt{n+1}\, \Psi_{m,n}, \qquad 
\mathbb{B}^{-}\Psi_{m,n}=e^{i\eta}\, \sqrt{m}\, \Psi_{m-1,n}, \quad n\neq0.
\end{equation}
However, when the eigenstate $n=0$ is involved, we get
\begin{equation*}
\mathbb{A}^{-}\Psi_{m,1}=\frac{1}{\sqrt{2}}\, e^{i\delta}\, \Psi_{m,0}, \qquad \mathbb{B}^{-}\Psi_{m,0}=\sqrt{m}\cos(\eta)\ \Psi_{m-1,0},
\end{equation*}
which spoils formulas (\ref{45a}) valid only for $n\neq0$. Therefore, we must complement formulas (\ref{45a}) with some others defined ``ad hoc" for $n=0$, so that they are all consistent, as follows
\begin{equation}\label{47a}
\mathbb{A}^{-}\Psi_{m,1}:=e^{i\delta}\, \Psi_{m,0}, \qquad \mathbb{B}^{-}\Psi_{m,0}:=\sqrt{m}\, e^{i\eta}\, \Psi_{m-1,0}.
\end{equation}
Once  $\mathbb{A}^{\pm}$ and $\mathbb{B}^{\pm}$ are defined in that way, these operators satisfy the following commutation relations (restricted to the subspace spanned by eigenstates):
 	\begin{equation}
	\left[\mathbb{A}^-,\mathbb{A}^+\right]=\mathbb{I}, \qquad
	\left[\mathbb{B}^-,\mathbb{B}^+\right]=\mathbb{I}, \qquad
	\left[\mathbb{A}^\pm,\mathbb{B}^\pm\right]=[\mathbb{A}^\pm,\mathbb{B}^\mp]= \mathbb{O}. \label{42c}
	\end{equation}
Since $\mathbb{A}^{-}$ and $\mathbb{B}^{-}$ commute, in a similar way to the scalar case~\cite{mm69,d17,dknn17}, we can build two-dimensional coherent states $\Upsilon_{\alpha,\beta}$ in graphene as the common eigenstates of both generalized annihilation operators,
\begin{equation}\label{43}
\mathbb{A}^- \Upsilon_{\alpha,\beta}=\alpha\, \Upsilon_{\alpha,\beta}, 
\qquad \mathbb{B}^- \Upsilon_{\alpha,\beta}=\beta\, \Upsilon_{\alpha,\beta}, 
\qquad \alpha,\beta\in\mathbb{C}.
\end{equation}
In general, these states will be superpositions of the eigenstates $\Psi_{m,n}$,
\begin{equation}\label{44}
\Upsilon_{\alpha,\beta}=\mathcal{N}_{\alpha,\beta}\sum_{m,n=0}^{\infty}c_n^\alpha \,d_m^\beta\, \Psi_{m,n}=\mathcal{N}_{\alpha,\beta} \sum_{m=0}^{\infty}d_m^\beta\, {\Pi_{m,\alpha}}=\mathcal{N}_{\alpha,\beta} \sum_{n=0}^{\infty}c_n^\alpha\, \Pi_{\beta,n},
\end{equation}
where $\mathcal{N}_{\alpha,\beta}$ are normalization constants. Taking specific sums over one of the quantum numbers, $n$ or $m$, we can construct the so-called partial coherent states $\Pi_{m,\alpha}$ and $\Pi_{\beta,n}$ \cite{mm69}, that fulfill the independent eigenvalue equations
\begin{equation}\label{45}
\mathbb{A}^-\, \Pi_{m,\alpha}=\alpha\, \Pi_{m,\alpha}, \qquad 
\mathbb{B}^-\, \Pi_{\beta,n}=\beta\, \Pi_{\beta,n}.
\end{equation}
In the remaining part of the present section we will explicitly build these two independent families of partial coherent states $\Pi_{m,\alpha}$ and $\Pi_{\beta,n}$ and, after that, the two-dimensional coherent states in graphene $\Upsilon_{\alpha,\beta}$ for some particular values of the parameters $\delta$ and $\eta$.

\subsection{Cyclotron motion}

	\begin{figure}[htb]
	\centering
	\includegraphics[width=.4\linewidth]{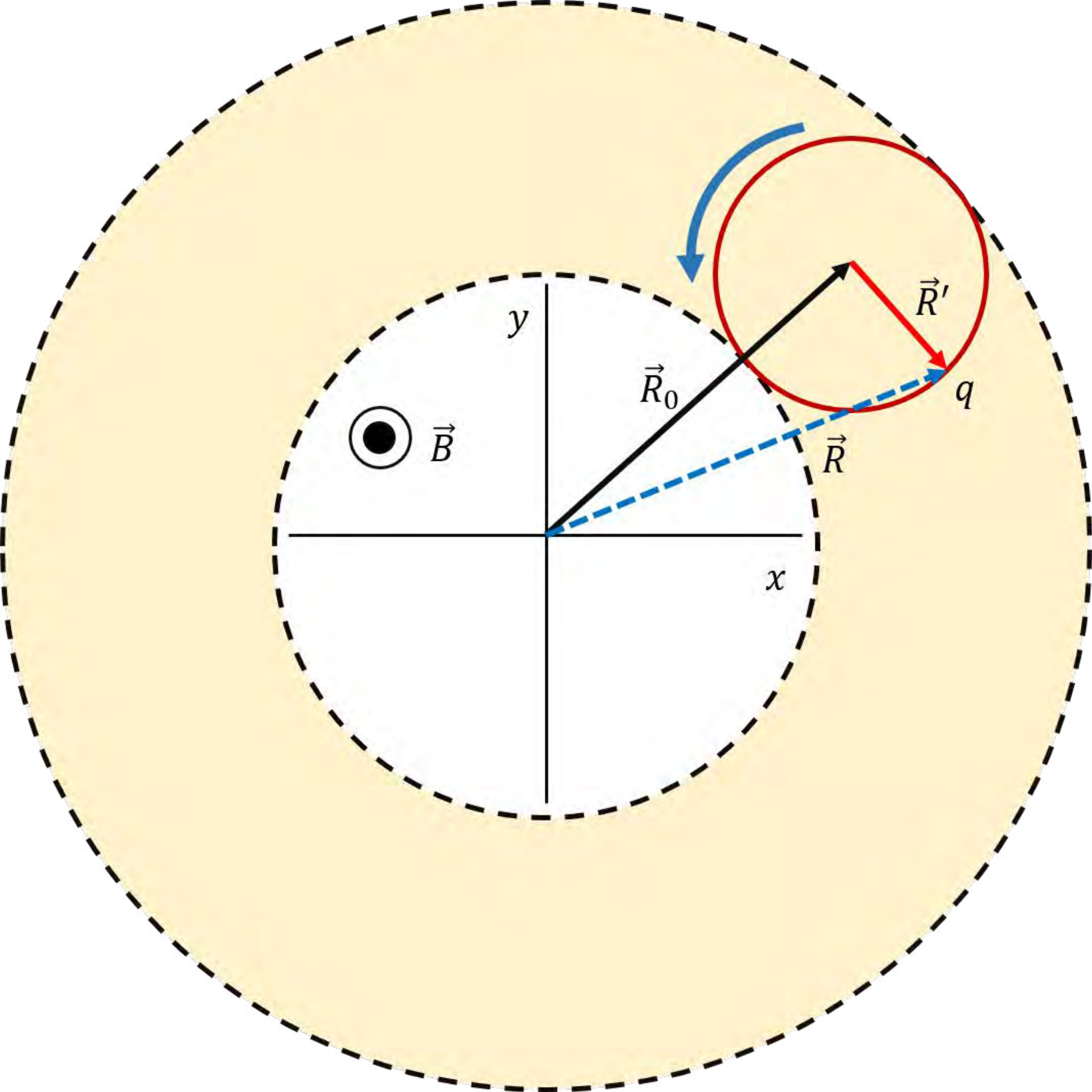}
	\caption{\label{fig:circle} Classical circular trajectory for an electron in a homogeneous magnetic field $\vec{B}$ perpendicular to the plane of the trajectory. Vector $\vec{R}_{0}$ locates the center of the orbit around which the particle moves,  while vector $\vec{R}\,'$ is the position vector of the particle with respect to the point $\vec{R}_{0}$.}
\end{figure}

In classical mechanics, due to the Lorentz force, a charged particle in a constant magnetic field follows a circular orbit whose radius is inversely proportional to the magnetic field strength (see Figure~\ref{fig:circle}). Now, to analyze the semi-classical motion through the coherent states defined above, let us consider the dimensionless {\it magnetic translation operators}, defined as \cite{z64,b64,l83}
\begin{equation}
	X_{0}=\frac{x}{\ell_{\rm B}}-\frac{\ell_{\rm B}}{\hbar}\left(p_y + \frac{\hbar}{2\ell_{\rm B}^{2}}x\right), 
	\quad 
	Y_{0}=\frac{y}{\ell_{\rm B}}+\frac{\ell_{\rm B}}{\hbar}\left(p_x - \frac{\hbar}{2\ell_{\rm B}^{2}}y\right),
\end{equation}
which can be expressed in terms of the operators $B^{\pm}$ as
\begin{equation}\label{magneticoperators}
	X_{0}   =\frac{1}{\sqrt{2}}(B^{-}+B^{+}), \quad 
	Y_{0}   =\frac{1}{\sqrt{2}i}(B^{-}-B^{+}).
\end{equation}

Analogously, we take into account the dimensionless position operators of a charged particle in a circular trajectory centered at the point $(X_{0},Y_{0})$, given by  
	\begin{equation}\label{positionoperators}
		R'_x=\frac{x}{\ell_{\rm B}} -X_{0}=\frac{1}{\sqrt{2}}(A^{-}+A^{+}), \qquad 
		R'_y=\frac{y}{\ell_{\rm B}}-Y_{0}=\frac{i}{\sqrt{2}}(A^{-}-A^{+}),
	\end{equation}
as well as the operator of the square of the distance from the center of the classical circular orbit to the origin of coordinates,	\begin{equation}\label{squaredistance}
		R_{0}^{2}=X_{0}^{2}+Y_{0}^{2}=2B^{+}B^{-}+1,
	\end{equation}
and the operator corresponding to the radius of the classical circular trajectory
	\begin{equation}\label{classicaltrajectory}
		(R')^{2}=(R'_x)^{2}+(R'_y)^{2}=2A^{+}A^{-}+1.
	\end{equation}
	
Now, to use a more compact notation in the next sections, we define the operators~\cite{diaz20}
	\begin{equation}\label{operatorsuv}
		u_{q}=\frac{1}{\sqrt{2}\, i^{q}}(B^{-}+(-1)^{q}B^{+}), \qquad v_{q}=\frac{i^{q}}{\sqrt{2}}(A^{-}+(-1)^{q}A^{+}), \qquad q=0,1,
	\end{equation}
such that
	\begin{equation}
		u_{0}\equiv X_{0}, \quad u_{1}\equiv Y_{0}, \qquad v_{0}\equiv R_{x}, \quad v_{1}\equiv R_{y}.
	\end{equation}
Hence, we can build the following matrix operators:
	\begin{equation}\label{matrixoperators}
		\mathcal{U}_{q}=u_{q}\otimes\mathbb{I}, \quad
		\mathcal{V}_{q}=v_{q}\otimes\mathbb{I}, \quad 
		\mathcal{R}_{0}^{2}=R_{0}^{2}\otimes\mathbb{I}, \quad (\mathcal{R'})^{2}=(R')^{2}\otimes\mathbb{I},
	\end{equation}
whose mean values will be calculated in the following subsections using the partial coherent states  $\Pi_{\beta,n}$ and 
$\Pi_{m,\alpha}$, and the two-dimensional coherent states $\Upsilon_{\alpha,\beta}$.

\subsection{First family of partial coherent states}\label{sec3.1}
Let us consider the operator $\mathbb{B}^-$ defined in eqs.~(\ref{41a})--(\ref{47a}), and consider the adjoint operator 
$\mathbb{B}^+$ given by
\begin{equation}\label{47}
 \mathbb{B}^+=\left[\begin{array}{c c}
\cos\eta\,B^+ & -\sin\eta\,\frac{B^+}{\sqrt{N+1}}A^- \\
\sin\eta\,A^+\frac{B^+}{\sqrt{N+1}} & \cos\eta\,B^+
\end{array}\right], \qquad \mathbb{B}^+\Psi_{m-1,n}=\sqrt{m}\,e^{-i\eta}\, \Psi_{m,n}.
\end{equation}
Then, the following commutation relations are fulfilled:
\begin{equation}\label{48}
[\mathbb{B}^-,\mathbb{B}^+]=\mathbb{I}, \qquad [H_{\rm DW},\mathbb{B}^\pm]=\mathbb{O}.
\end{equation}
The first family of partial coherent states is composed by the pseudo-spinor states $\Pi_{\beta,n}$  that  satisfy the following equations:
	\begin{eqnarray}
\mathbb{B}^- \Pi_{\beta,n}&=&\beta\, \Pi_{\beta,n}, \quad \beta\in\mathbb{C}, \label{49a} \\
H_{\rm DW} \Pi_{\beta,n}&=&\hbar\,\omega\,\sqrt{n}\,\Pi_{\beta,n}, \quad n=0,1,2,\dots, \label{49b}
\end{eqnarray}
where
\begin{equation}\label{50}
\Pi_{\beta,n}=(1-\delta_{0n})\sum_{m=0}^{n-1}c_{m,n}\Psi_{m,n}^{+}+\sum_{m=n}^{\infty}d_{m,n}\Psi_{m,n}^{-}, \quad n=0,1,2,\dots
\end{equation}
Therefore, when substituting in the eigenvalue equation, the partial coherent states with a well-defined energy $E_n=\sqrt{n}\,\hbar\,\omega$ turn out to be
\begin{equation}\label{51}
\Pi_{\beta,n}=\exp\left(-\frac{\vert\tilde{\beta}\vert^2}{2}\right)\left((1-\delta_{0n})\sum_{m=0}^{n-1}\frac{\tilde{\beta}^m}{\sqrt{m!}}\Psi_{m,n}^{+}+\sum_{m=n}^{\infty}\frac{\tilde{\beta}^m}{\sqrt{m!}}\Psi_{m,n}^{-}\right),
\end{equation}
where $\tilde{\beta}\equiv\exp\left(-i\eta\right)\beta$. The parameter $\eta$ can be considered as an additional phase for the eigenvalue $\beta$. It is possible to identify the up or down scalar coherent states of the operator $\mathbb{B}^- $ for each energy level $n$ as
	\begin{align*}
\psi^u_{\beta,n-1}&=  e^{-\vert\tilde{\beta}\vert^2 /2}\, \left( \sum_{m=0}^{n-1}\frac{\tilde{\beta}^m}{\sqrt{m!}}\, \psi_{m,n-1}^{+}+\sum_{m=n}^{\infty}\frac{\tilde{\beta}^m}{\sqrt{m!}}\, \psi_{m,n-1}^{-}\right), 
\\
\psi^d_{\beta,n}&=e^{-\vert\tilde{\beta}\vert^2 /2}\, \left((1-\delta_{0n})\sum_{m=0}^{n-1}\frac{\tilde{\beta}^m}{\sqrt{m!}}\, \psi_{m,n}^{+}+\sum_{m=n}^{\infty}\frac{\tilde{\beta}^m}{\sqrt{m!}}\, \psi_{m,n}^{-}\right). 
\end{align*}
Hence, the partial pseudo-spinor coherent states of Eq.~(\ref{51}) can be expressed as
\begin{equation}\label{53}
\Pi_{\beta,n}=\frac{1}{\sqrt{2^{(1-\delta_{0n})}}}\left(\begin{array}{c}
(1-\delta_{0n})\psi^u_{\beta,n-1} \\ [1ex]
i\psi^d_{\beta,n}
\end{array}\right), \quad n=0,1,2,\dots
\end{equation}

\subsubsection{Displacement operator}\label{sec3.1.1}

In this subsection, we will see how the coherent states $\Pi_{\beta,n}$ are also obtained by acting with an unitary operator identified as a displacement operator, on the pseudo-spinor states $\Psi_{0,n}$ whose scalar components have the maximum value of angular momentum in $z$-direction $l=n$ ($m=0$ in Figure~\ref{fig:diagram}). Such a set of states satisfy:
	\begin{equation}
	(\mathbb{B}^+)^k\, \Psi_{0,n}^+=\sqrt{k!}\,e^{i\eta k}\, \Psi_{k,n}^{\pm}. \label{54}
	\end{equation}
Considering the displacement operator $\mathbb{D}(\lambda)$ given by
\begin{equation}\label{55}
\mathbb{D}(\lambda)=\exp\left(\lambda\mathbb{B}^+-\lambda^\ast\mathbb{B}^-\right)=e^{-\vert\lambda\vert^2/2}\, \exp\left(\lambda\mathbb{B}^+\right)\exp\left(-\lambda^\ast\mathbb{B}^-\right),
\end{equation}
acting on the states $\Psi_{0,n}^+$, we find that
\begin{eqnarray}\label{56}
\nonumber \mathbb{D}(\lambda)\Psi_{0,n}^+&=& e^{-\vert\lambda\vert^2/2}\, \exp\left(\lambda\mathbb{B}^+\right)\exp\left(-\lambda^\ast\mathbb{B}^-\right)\Psi_{0,n}^+ =e^{-\vert\lambda\vert^2/2}  \left( (1-\delta_{0n})\sum_{m=0}^{n-1}\frac{\tilde{\lambda}^m}{\sqrt{m!}}\Psi_{m,n}^+ +\sum_{m=n}^{\infty}\frac{\tilde{\lambda}^m}{\sqrt{m!}}\Psi_{m,n}^- \right),
\end{eqnarray}
where the Baker-Campbell-Haussdorff relation has been employed and $\tilde{\lambda}=\lambda\,\exp\left(-i\eta\right)$. Up to a normalization factor, this expression coincides with that of Eq.~(\ref{51}) if $\tilde{\beta}=\tilde{\lambda}$. In particular, taking $\eta=2k\pi$, $k=0,1,\dots$, we have $\tilde{\lambda}=\tilde{\beta}=\beta$, and in this case the partial coherent states $\Psi_{\beta,n}$ can be rewritten as
\begin{equation}\label{57}
\Pi_{\beta,n}=\mathbb{D}(\beta)\Psi_{0,n}^+=\exp\left(\beta\mathbb{B}^+-\beta^\ast\mathbb{B}^-\right)\Psi_{0,n}^+, \quad n=0,1,2,\dots,
\end{equation}
with
\begin{equation}\label{58}
\mathbb{B}^-=\left[\begin{array}{c c}
B^- & 0 \\
0 & B^-
\end{array}\right].
\end{equation}

Finally, to give an analytical expression for the scalar coherent states $\psi^u_{\beta,n-1}$ and $\psi^d_{\beta,n}$ in Eq.~(\ref{53}) for $n\neq0$, we define the complex variable $z$ as (see Malkin-Man'ko~\cite{mm69})
\begin{equation}\label{59}
z=\xi\exp(i\theta)=\frac{\sqrt{2}}{\ell_{\rm B}}\left(\frac{x+iy}{2}\right),
\end{equation}
and therefore the operators $A^\pm$ and $B^\pm$ in eqs.~(\ref{15}) and (\ref{35}) can be rewritten as
	\begin{equation}
	A^-=\partial_{z}+\frac{z^\ast}{2}, \quad A^+=-\partial_{z^\ast}+\frac{z}{2}, \qquad
	B^-=\partial_{z^\ast}+\frac{z}{2}, \quad B^+=-\partial_{z}+\frac{z^\ast}{2}. \label{60b}
	\end{equation}
The action of the annihilation operator $\mathbb{B}^-$ in~(\ref{58}) on the states $\Pi_{\beta,n}$ in~(\ref{53}), gives the following expressions for each component of the pseudo-spinor:
	\begin{eqnarray}
	B^-\psi^u_{\beta,n-1}&=\beta \psi^u_{\beta,n-1} \quad\Rightarrow\quad \psi^u_{\beta,n-1}&=\exp\left(\left(\beta-\frac{z}{2}\right)z^\ast\right)g_n(z), \label{61b}
	\\	B^-\psi^d_{\beta,n}&=\beta \psi^d_{\beta,n} \qquad\Rightarrow\qquad \psi^d_{\beta,n}&=\exp\left(\left( \beta-\frac{z}{2}\right) z^\ast\right)f_n(z), \label{61a}
	\end{eqnarray}
where $f_n(z)$ y $g_n(z)$ are functions to be determined. Next, according to Eq.~(\ref{49b}), each component of 
$\Pi_{\beta,n}$ satisfies, respectively,
	\begin{eqnarray}
	\mathcal{H}_1 \psi^u_{\beta,n-1}&=n\,\psi_{\beta,n-1} \Rightarrow\quad  (z-\beta)\frac{dg_n(z)}{dz}&=(n-1)\,g_n(z), \label{62a}\\
	\mathcal{H}_2 \psi^d_{\beta,n}& =n\,\psi_{\beta,n} \quad\Rightarrow\quad (z-\beta)\frac{df_n(z)}{dz}&=n\,f_n(z), \label{62b}
	\end{eqnarray}
whose solutions are, in each case,
\begin{equation}\label{63}
f_n(z)=f_{0}(z-\beta)^n, \quad g_n(z)=g_{0}(z-\beta)^{n-1},
\end{equation}
where $f_{0}$, $g_{0}$ are constants to be fixed. Finally, after replacing in (\ref{49b}), we get that $g_{0}=\sqrt{n}f_{0}$ and then the normalized pseudo-spinor partial coherent states $ \Pi_{\beta,n}$, $n=0,1,2,\dots,$ are given by (setting $\exp\left(i\eta\right)=1$)
\begin{equation}\label{64}
\Pi_{\beta,n}(x,y)=\frac{1}{\sqrt{2^{(1-\delta_{0n})}\pi\,n!}}\exp\left(\left( \beta-\frac{z}{2}\right) z^\ast-\frac{\vert\beta\vert^2}{2}\right)
\left(
\begin{array}{c}
\sqrt{n}(z-\beta)^{n-1}\\
i(z-\beta)^{n}
\end{array}
\right).
\end{equation}

\subsubsection{Probability and current densities}\label{sec3.1.2}
\begin{figure}[htb]
	\centering
	\begin{minipage}[b]{0.33\textwidth}
		\includegraphics[width=\textwidth]{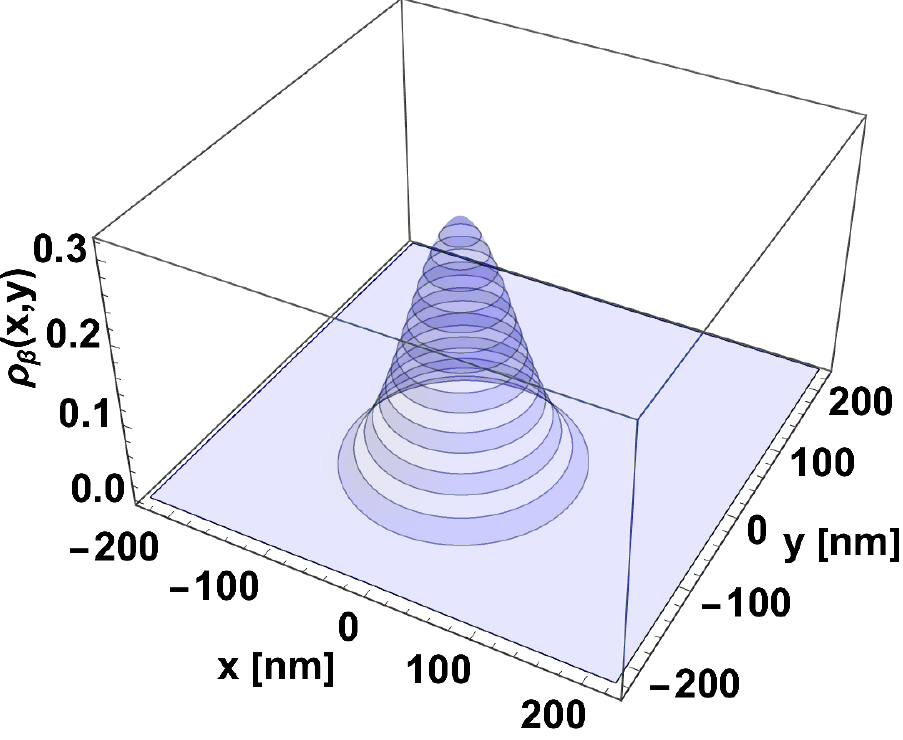}\\
		\centering{\footnotesize (a) $\rho_{\beta,0}$ with $\beta=\exp(-i\pi/2)$.}
		\label{fig:rho_beta0_i}
	\end{minipage}
\hspace{3cm}
	~ 
	\begin{minipage}[b]{0.33\textwidth}
		\includegraphics[width=\textwidth]{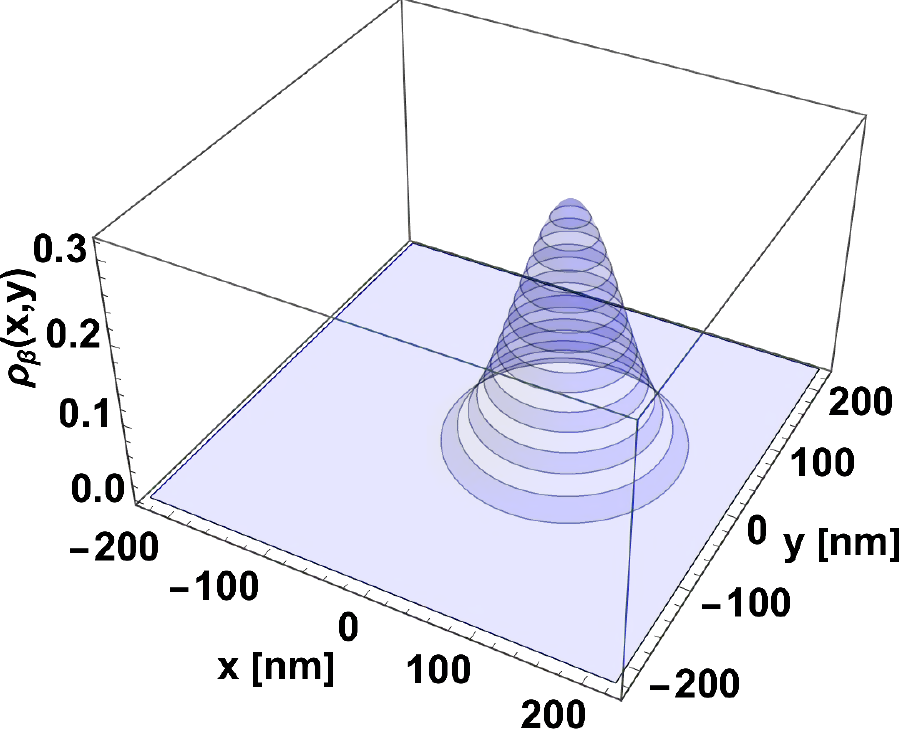}\\
		\centering{\footnotesize (b) $\rho_{\beta,0}$ with $\beta=1$.}
		\label{fig:rho_beta0_ii}
	\end{minipage}
	\\ [2ex]
	~ 
	\begin{minipage}[b]{0.33\textwidth}
		\includegraphics[width=\textwidth]{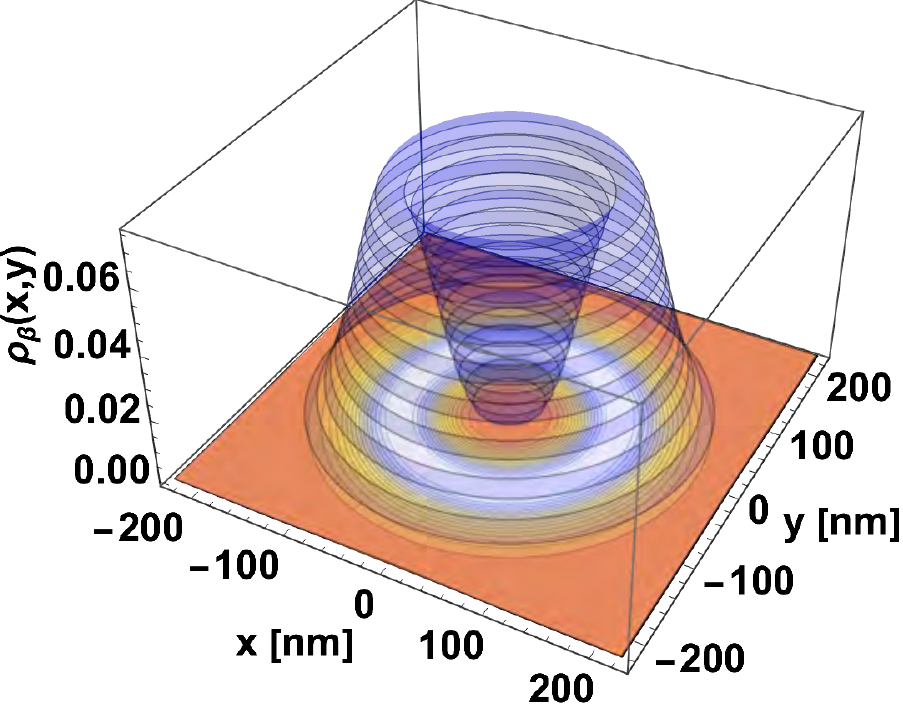}\\
		\centering{\footnotesize (c) $\beta=1$ and $n=3$.}
		\label{fig:j_beta1_i}
	\end{minipage}
\hspace{3cm}
	~ 
	\begin{minipage}[b]{0.33\textwidth}
		\includegraphics[width=\textwidth]{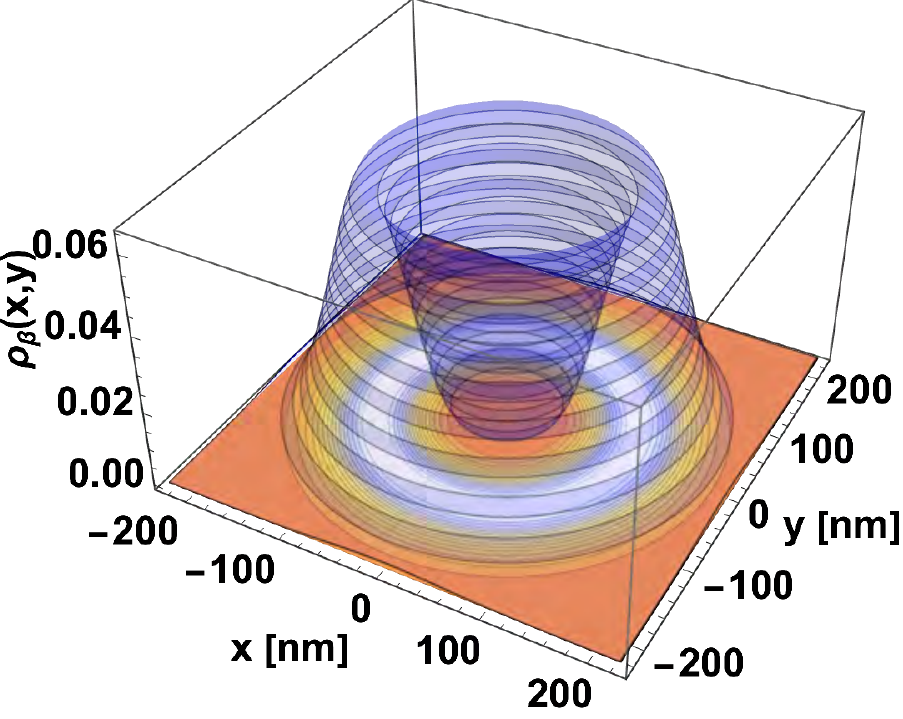}\\
		\centering{\footnotesize (d) $\beta=1$ and $n=4$.}
		\label{fig:j_beta1_ii}
	\end{minipage}
	\caption{\label{fig:rho_j_beta1}In (a) and (b) the probability density $\rho_{\beta,0}(x,y)$ from \eqref{65} is shown for $\beta=\vert\beta\vert\exp(i\varphi)$ with $B_{0}=0.3$ T. In (c) and (d), for the partial coherent states $\Pi_{\beta,n}$ with $\beta=1$, the probability density $\rho_{\beta,n}$ is shown in the 3D plots, and the angular current density $j_{\beta,n,\vec{u}_\theta}/(e\,v_{\rm F})$ from \eqref{69b} is shown in the 2D plots ($x-y$ plane), for $n=3,4$.}
\end{figure}

The probability density $\rho_{n,\beta}(x,y)$ for the partial coherent states $\Pi_{\beta,n}$ in \eqref{64} is given by
\begin{equation}\label{65}
\rho_{\beta,n}(x,y)=\Pi^\dagger_{\beta,n} \Pi_{\beta,n}=\frac{e^{-\vert z-\beta\vert^2}}{2^{(1-\delta_{0n})}\pi\,n!}\vert z-\beta\vert^{2n-2}(\vert z-\beta\vert^{2}+n),
\end{equation}
where $\beta=\vert\beta\vert\exp(i\varphi)$, $n=0,1,2,\dots$, and
	\begin{equation}
 	\vert z-\beta \vert^2=\xi^2+\vert\beta\vert^2-2\xi\vert\beta\vert\cos(\theta-\varphi)=\frac{2}{\ell_{\rm B}^{2}}\frac{x^2+y^2}{4}+\vert\beta\vert^2-\frac{\sqrt{2}}{\ell_{\rm B}}\vert\beta\vert\left(x\cos\varphi+y\sin\varphi\right). \label{65b}
	\end{equation}
Some examples of probability density $\rho_{n,\beta}(x,y)$ for partial coherent states $\Pi_{\beta,n}$ are shown in Figure~\ref{fig:rho_j_beta1}, where it is evident that the coherent states $\Pi_{\beta,n}(x,y)$ are displaced from the origin, similarly to the standard coherent states. They are centered around the point $(x_0,y_0)$, 
\begin{equation}\label{66}
(x_0,y_0)=\sqrt{2}\vert\beta\vert\ell_{\rm B}\left(\cos\varphi,\sin\varphi\right),
\end{equation}
which obviously depends on $\beta$, and represents, in a classical interpretation, the center of a circle in the 
$x-y$ plane along which the classical particle is moving under the action of the magnetic field~\cite{kr05}.

If $(x',y')=(r'\cos\theta',r'\sin\theta')$ denotes the coordinates of a point with respect to a reference frame centered at $(x_0,y_0)$, then the coordinates of the point with respect to a frame whose center is $(0,0)$ are
\begin{equation}\label{67}
x=\sqrt{2}\vert\beta\vert\ell_{\rm B}\cos\varphi+r'\cos\theta', \quad y=\sqrt{2}\vert\beta\vert\ell_{\rm B}\sin\varphi+r'\sin\theta'.
\end{equation}
Hence,
\begin{equation}\label{68}
  z=\frac{\sqrt{2}}{\ell_{\rm B}}\,\frac{x+iy}{2}=(\vert\beta\vert\cos\varphi+\xi'\cos\theta')+i(\vert\beta\vert\sin\varphi+\xi'\sin\theta')=\beta+z',
\end{equation}
where $\xi'=r'/(\sqrt{2}\ell_{\rm B})$ and $z'=\xi'\exp(i\theta')$.

Thus, the current densities  $j_{\beta,n,\vec{u}}$ for $n\neq0$ of the partial coherent states $\Pi_{\beta,n}$ along to the directions of the unit vectors $\vec{u}_{\xi'}$ and $\vec{u}_{\theta'}$ in the displaced frame are 
\begin{eqnarray}
j_{\beta,n,\vec{u}_{\xi'}}(\xi')\!&\!=\!& \!
ev_{\rm F} \, \Pi^\dagger_{\beta,n}(\vec{\sigma}\cdot\vec{u}_{\xi'})\, \Pi_{\beta,n}=0, \label{69a} 
\\ 
 j_{\beta,n,\vec{u}_{\theta'}}(\xi')\!&\!=\!&\! ev_{\rm F}\,  \Pi^\dagger_{\beta,n} (\vec{\sigma}\cdot\vec{u}_{\theta'})\,\Pi_{\beta,n} =\frac{2\,ev_{\rm F}\sqrt{n}}{2^{(1-\delta_{0n})}\pi\,n!}(\xi')^{2n-1} e^{-\xi'^2}. \label{69b}
\end{eqnarray}
Again it  is evident that there is no probability flux in the radial direction $\xi'$, as it is expected due to the symmetry of the problem. It is also evident that as $n$ increases, the probability amplitude decreases, while the minimum value of the angular current density moves away radially from the origin, as can be seen in Figures~\ref{fig:rho_j_beta1}(a) and \ref{fig:rho_j_beta1}(b). We observe that the probability density of the partial coherent states for $n=0$ has a Gaussian distribution while for $n\neq0$ does not. This is due essentially to the fact that these partial coherent states are obtained by magnetic translational operators acting on the ground state $\Psi_{0,n}$. In a classical interpretation,  electrons rotate around a point $(x_0,y_0)$, located at a distance $d=\sqrt{2}\vert \beta\vert\ell_{\rm B}$ from the origin; as their energy $E_n$ increases, they are located further away from such a center. These features can be appreciated in the examples shown in 
Figures~\ref{fig:rho_j_beta1}(a)--\ref{fig:rho_j_beta1}(d).

\subsubsection{Cyclotron motion}

After a straightforward calculation, the mean values of the matrix operators in 
eqs.~\eqref{squaredistance}--(\ref{matrixoperators}) for the partial coherent states $\Pi_{\beta,n}$ obtained in \eqref{64} turn out to be
	\begin{eqnarray}
		\langle\mathcal{U}_{q}\rangle_{\beta,n}=\frac{\beta+(-1)^{q}\beta^{\ast}}{\sqrt{2}\, i^{q}}, \qquad \langle\mathcal{V}_{q}\rangle_{\beta,n}=0, \qquad
		\langle\mathcal{R}_{0}^{2}\rangle_{\beta,n}=2\vert\beta\vert^{2}+1, \qquad 
		\langle(\mathcal{R}')^{2}\rangle_{\beta,n}=\begin{cases}
			1, & n=0, \\
		2n, & n\neq0.
		\end{cases}
	\end{eqnarray}
The results of $\ell_{\rm B}\langle\mathcal{U}_{q}\rangle_{\beta,n}$ and 
$\ell_{\rm B}^{2} \langle(\mathcal{R}')^{2}\rangle_{\beta,n}$ agree with Eq.~(\ref{66}) and those in~\cite{diaz20}, respectively. The latter also corresponds to the mean value of $(\mathcal{R}')^{2}$ for the eigenstates $\Psi_{m,n}^{\pm}$, since the partial coherent states $\Pi_{\beta,n}$ are basically equal to the pseudo-spinor eigenstates $\Psi_{m,n}^{\pm}$ but centered on the point $(x_{0},y_{0})$. In addition, according to $\langle\mathcal{R}_{0}^{2}\rangle_{\beta,n}$, as $\vert\beta\vert$ increases, the center of the classical trajectory moves away from the coordinate origin (see Fig.~\ref{fig:circle}).

\subsection{Second family of partial coherent states}\label{sec3.2}
Now, let us consider the operator $\mathbb{A}^-$ defined in eqs.~(\ref{41a})--(\ref{47a}), such that
\begin{equation}\label{70}
\mathbb{A}^-\Psi_{m,n}=\exp(i\delta)\sqrt{n}\,\Psi_{m,n-1}, \quad n=0,1,2,\dots
\end{equation}
This operator is related with one of the annihilation operators in \cite{df17} for $\delta=0$ within the nonlinear algebras formalism \cite{mmsz93,mmsz93a,hh02,rr00,rr00a,s00}.
One can construct the second family of partial coherent states, associated with the operator $\mathbb{A}^-$ as the pseudo-spinor states $\Pi_{m,\alpha}$ such that
	\begin{eqnarray}
\mathbb{A}^- \Pi_{m,\alpha}=\alpha\, \Pi_{m,\alpha}, \quad \alpha\in\mathbb{C}, 
\qquad
\mathbb{B}^+\mathbb{B}^- \Pi_{m,\alpha}=m\, \Pi_{m,\alpha}, \quad m=0,1,2,\dots, 
\end{eqnarray}
where
\begin{equation}\label{77}
\Pi_{m,\alpha}=\sum_{n=0}^{m}c_{m,n}\Psi_{m,n}^{-}+\sum_{n=m+1}^{\infty}d_{m,n}\Psi_{m,n}^{+}, \quad m=0,1,2,\dots,
\end{equation}
the pseudo-spinor states $\Psi_{m,n}^{\pm}$ given by \eqref{34aaa}--\eqref{34bbb}.
By applying the eigenvalue equation that defines the partial coherent states, the states $\Pi_{m,\alpha}$ turn out to be
\begin{eqnarray}\label{78}
\Pi_{m,\alpha}=\frac{1}{\sqrt{2 e^{\vert\tilde{\alpha}\vert^2} -1}}\left( \Psi_{m,0}^- +(1-\delta_{0m})\sum_{n=1}^{m}\frac{\sqrt{2}\tilde{\alpha}^n}{\sqrt{n!}}\Psi_{m,n}^{-} 
+\sum_{n=m+1}^{\infty}\frac{\sqrt{2}\tilde{\alpha}^n}{\sqrt{n!}}\Psi_{m,n}^{+}\right),  
\end{eqnarray}
where $\tilde{\alpha}=\alpha e^{-i\delta}$ and $m=0,1,2,\dots$ The effect of $\delta$ is a phase change in $\alpha$, just as it happened with $\eta$ and $\beta$ before.

The coherent states $\Pi_{m,\alpha}$ present some important differences with respect those of the previous subsection $\Pi_{\beta,n}$. Due to the fact that the definition (\ref{41a}) does not allow $\mathbb{A}^{-}$ to be expressed as a pure differential operator, even for $\delta=0$ (it includes square roots of a number operator), the wave functions of the coherent states have no closed analytical expressions. In the same way, the interpretation of these coherent states as displaced wave functions, in the Perelomov approach~\cite{p72}, can not be fully implemented. These details imply that some features of resulting coherent states remain rather diffuse, as it will be shown in the sequel.

\subsubsection{Probability and current densities, and mean energy}\label{sec3.2.3}

\begin{figure}[htb]
	\centering
	\begin{minipage}[b]{0.34\linewidth}
		\includegraphics[width=\textwidth]{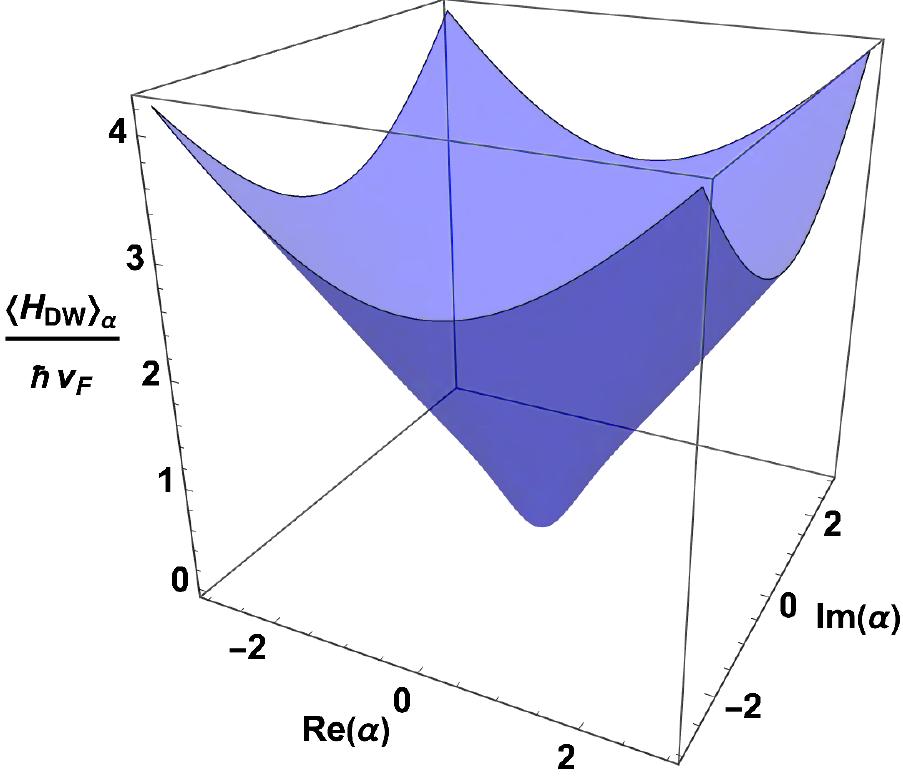}
		\label{fig:H_alpha1}
	\end{minipage}
	\caption{\label{fig:H_alpha}Mean energy $\langle H_{\rm DW}\rangle_{\alpha}/(\hbar\,v_{\rm F})$ with $B_{0}=0.3$ T as a function of $\alpha$, as given in (\ref{82}).}
\end{figure}

In the first place, it is not difficult to show that the mean value of the energy in the coherent state $\Pi_{m,\alpha}$ in (\ref{78}) is given by
\begin{equation}\label{82}
	\langle H_{\rm DW}\rangle_\alpha=\frac{2\hbar\,\omega}{2\exp\left(\vert\alpha\vert^2\right)-1}\sum_{n=0}^{\infty}\frac{\vert\alpha\vert^{2n}}{n!}\sqrt{n}.
\end{equation}
The mean energy of these coherent states, behaves as a continuous function of the eigenvalue $\alpha$ as $\langle H_{\rm DW}\rangle_\alpha\approx\vert\alpha\vert$, in agreement with the Hamiltonian form (\ref{5}) in terms of $A^{\pm}$.
A plot of this function is shown in Figure \ref{fig:H_alpha}.

\begin{figure}[htb]
	\centering
	\begin{minipage}[b]{0.33\textwidth}
		\includegraphics[width=\textwidth]{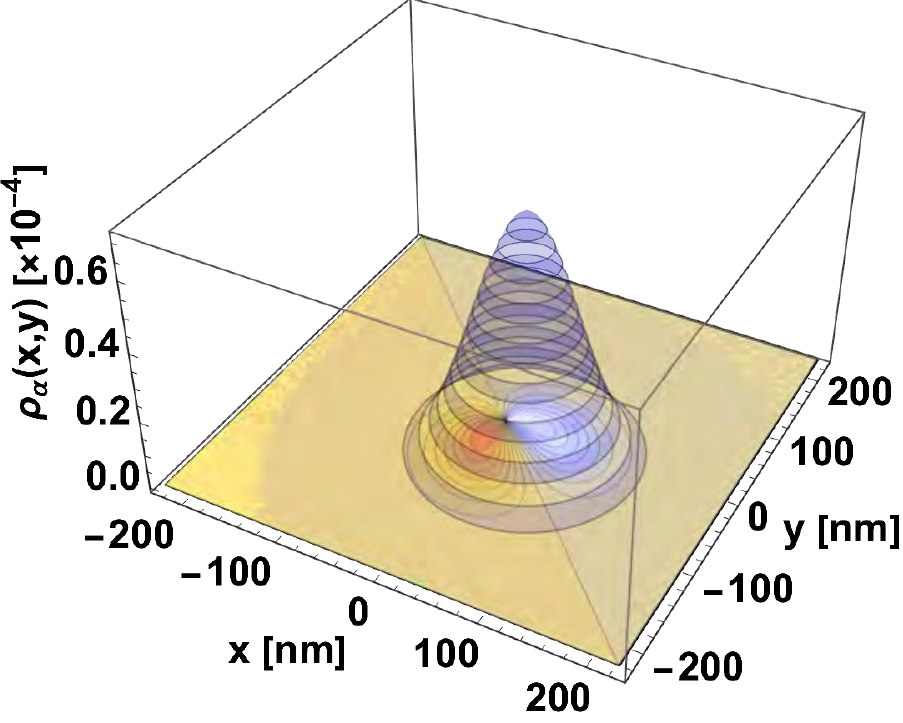}\\
		\centering{\footnotesize (a) $m=0$ and $\alpha=\exp(i\pi/2)$.}
		\label{fig:rho_alpha1_i}
	\end{minipage}
\hspace{3cm}
	\begin{minipage}[b]{0.33\textwidth}
		\includegraphics[width=\textwidth]{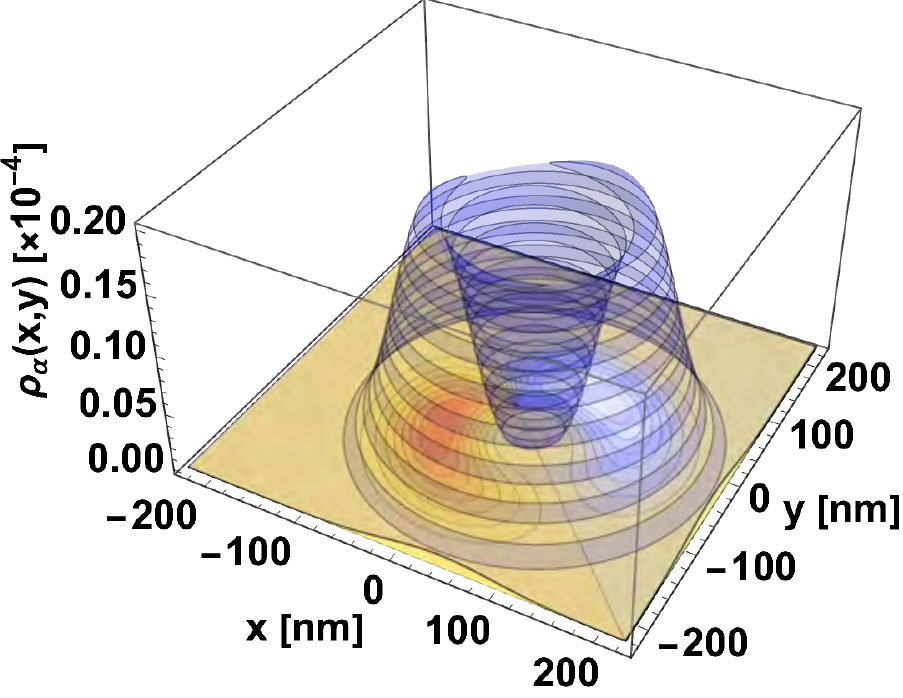}\\
		\centering{\footnotesize (b) $m=2$ and $\alpha=\exp(i\pi/2)$.}
		\label{fig:jrho_alpha1_i}
	\end{minipage}
\\ [2ex]
		\begin{minipage}[b]{0.33\textwidth}
		\includegraphics[width=\textwidth]{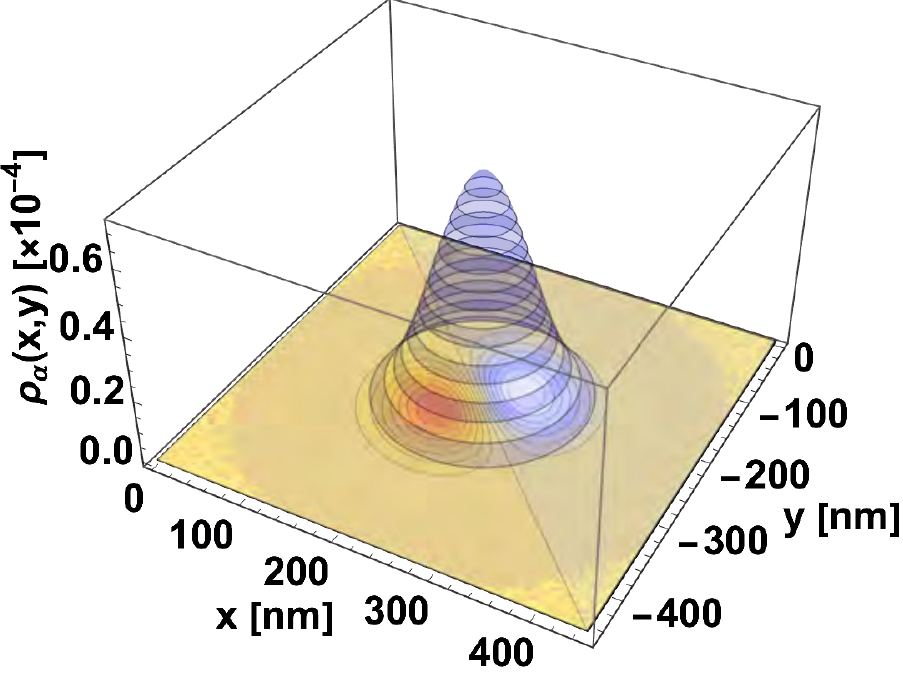}\\
		\centering{\footnotesize (c)  $m=0$ and $\alpha=5\exp(i\pi/2)$.}
		\label{fig:jtheta_alpha1_i}
	\end{minipage}
\hspace{3cm}
\begin{minipage}[b]{0.33\textwidth}
	\includegraphics[width=\textwidth]{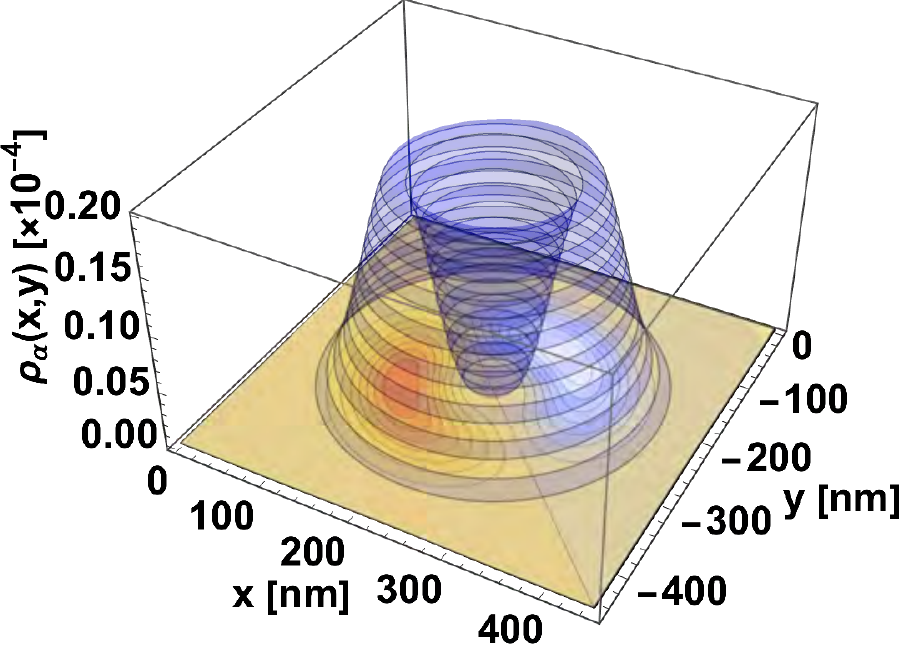}\\
	\centering{\footnotesize (d) $m=2$ and $\alpha=5\exp(i\pi/2)$.}
	\label{fig:jtheta_alpha1_i}
\end{minipage}
	\caption{\label{fig:rho_alpha1} Probability density $\rho_{m,\alpha}$ (the 3D plots) and radial current density $j_{m,\alpha,\vec{u}_{\xi}}/(e\,v_{\rm F})$ (the 2D plots in the $x-y$ plane) with $B_{0}=0.3$ T for some of the partial coherent states $\Pi_{m,\alpha}$ given in~(\ref{78}).}
\end{figure}

\begin{figure}[htb]
	\centering
	\begin{minipage}[b]{0.33\textwidth}
		\includegraphics[width=\textwidth]{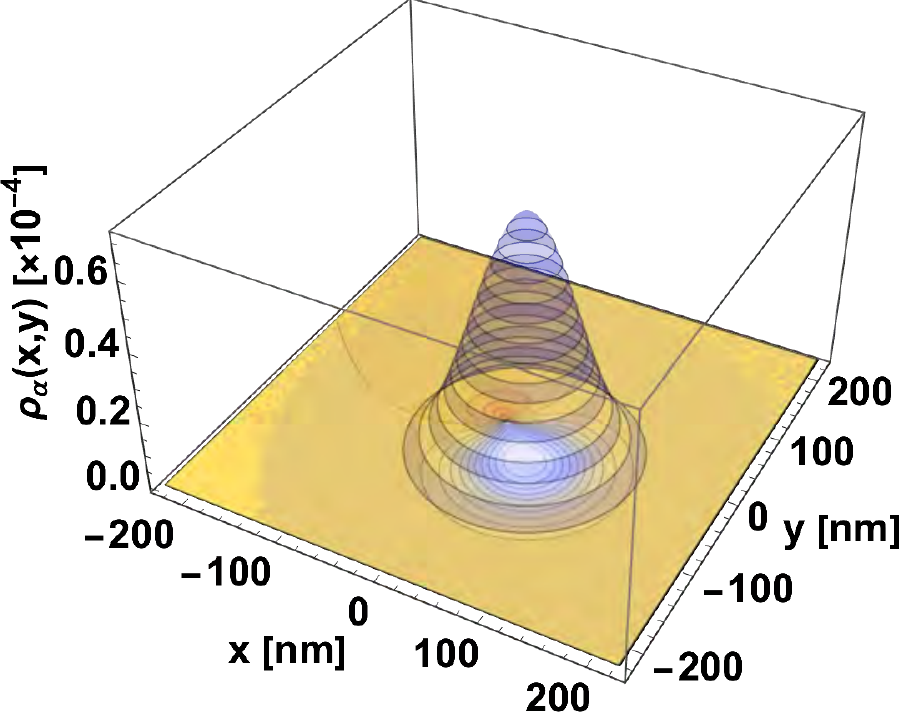}\\
		\centering{\footnotesize (a) $m=0$ and $\alpha=\exp(i\pi/2)$, }
		\label{fig:rho_alpha1_ii}
	\end{minipage}
\hspace{3cm}
	\begin{minipage}[b]{0.33\textwidth}
		\includegraphics[width=\textwidth]{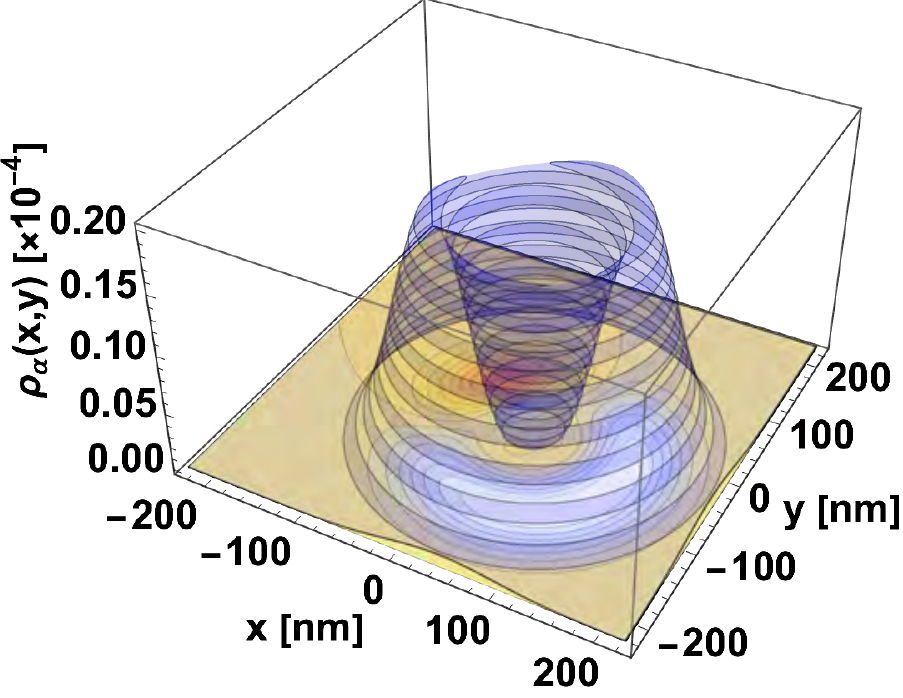}\\
		\centering{\footnotesize (b) $m=2$ and $\alpha=\exp(i\pi/2)$}
		\label{fig:jrho_alpha1_ii}
	\end{minipage}
\\ [2ex]
	\begin{minipage}[b]{0.33\textwidth}
		\includegraphics[width=\textwidth]{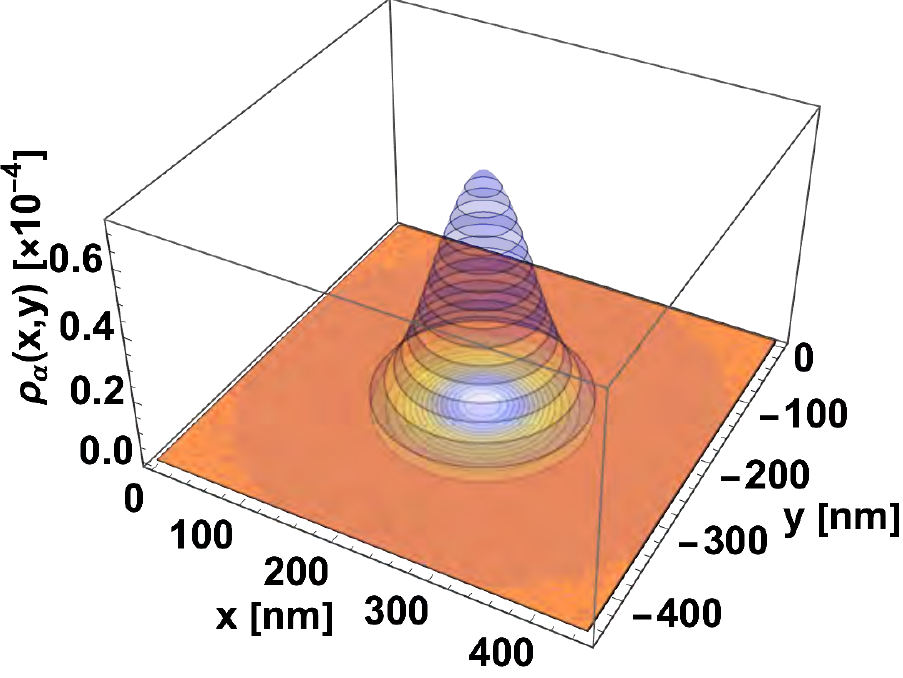}\\
		\centering{\footnotesize (c) $m=0$ and $\alpha=5\exp(i\pi/2)$}
		\label{fig:jtheta_alpha1_ii}
	\end{minipage}
\hspace{3cm}
\begin{minipage}[b]{0.33\textwidth} 
	\includegraphics[width=\textwidth]{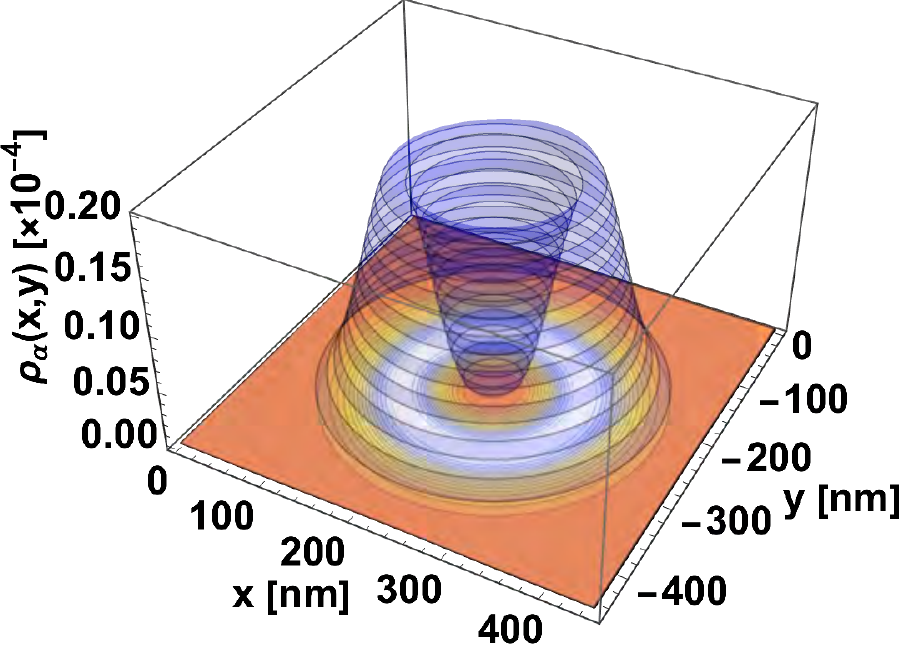}\\
	\centering{\footnotesize (d) $m=2$ and $\alpha=5\exp(i\pi/2)$}
	\label{fig:jtheta_alpha1_ii}
\end{minipage}
	\caption{\label{fig:rho_alpha2} Probability density $\rho_{m,\alpha}$ (the 3D plots) and angular current density $j_{m,\alpha ,\vec{u}_{\theta}}/(e\,v_{\rm F})$ (the 2D plots in the $x-y$ plane) with $B_{0}=0.3$ T for some of the coherent states $\Psi_{m,\alpha}$ given in~(\ref{78}).}
\end{figure}

To obtain expressions for the probability $\rho_{m,\alpha}$ and current $j_{m,\alpha,\vec{u}}$ densities of the coherent states in Eq.~(\ref{78}), the matrix operator $(\vec{\sigma}\cdot\vec{u})_k$ is defined as
\begin{equation}\label{79}
(\vec{\sigma}\cdot\vec{u})_k=\left[\begin{array}{c c}
0 & (-i)^{k}e^{-i\theta} \\
i^{k}e^{i\theta} & 0
\end{array}\right], \quad k=0,1, 
\end{equation}
such that $(\vec{\sigma}\cdot\vec{u})_0=\vec{\sigma}\cdot\vec{u}_{\xi}$ and $(\vec{\sigma}\cdot\vec{u})_1=\vec{\sigma}\cdot \vec{u}_{\theta}$. The expressions for the densities $\rho_{m,\alpha}$ and $j_{m,\alpha,\vec{u}}$ are straightforwardly computed but, as they have cumbersome expressions, we have moved them to Appendix~\ref{Appendix}. Some graphics of these densities are shown in Figures~\ref{fig:rho_alpha1} and \ref{fig:rho_alpha2}. As in the previous partial coherent states, the corresponding eigenvalue $\alpha$ indicates where the probability density is displaced in the $x-y$ plane, although without a clear point of location, as it happens for the coherent states $\Pi_{\beta,n}$ in Eq.~(\ref{64}). The value of $m$ modifies the shape of the probability distribution.

\subsubsection{Cyclotron motion}

By direct calculation we can prove that the mean values of the matrix operators in (\ref{matrixoperators}) for the partial coherent states $\Pi_{m,\alpha}$ are 
\begin{eqnarray}
	\langle\mathcal{U}_{q}\rangle_{m,\alpha}&=0 \qquad\qquad\quad \langle\mathcal{V}_{q}\rangle_{m,\alpha}&=\frac{i^{q}(\tilde{\alpha}+(-1)^{q}\tilde{\alpha}^{\ast})}{\sqrt{2}(2\exp\left(\vert\tilde{\alpha}\vert^2\right)-1)}\left(\exp\left(\vert\tilde{\alpha}\vert^{2}\right)+\sum_{n=1}^{\infty}\frac{\vert\tilde{\alpha}\vert^{2n}}{\sqrt{(n-1)!(n+1)!}}\right), \\ [1ex]
	\langle\mathcal{R}_{0}^{2}\rangle_{m,\alpha}&=2m+1, \quad \langle(\mathcal{R}')^{2}\rangle_{m,\alpha}&=\frac{4\vert\tilde{\alpha}\vert^{2}\exp\left(\vert\tilde{\alpha}\vert^{2}\right)+1}{2\exp\left(\vert\tilde{\alpha}\vert^2\right)-1}.
\end{eqnarray}
The results of $\langle\mathcal{V}_{q}\rangle_{m,\alpha}$ and $\langle(\mathcal{R}')^{2}\rangle_{m,\alpha}$ agree with those in~\cite{df17}, which correspond to a description by using a Landau-like gauge. Therefore, the partial coherent states $\Pi_{m,\alpha}$ describe the classical motion of the charged particle around a given point $(x_{0},y_{0})$ (see Fig.~\ref{fig:circle}).

\subsection{Two-dimensional coherent states}\label{sec3.3}

\begin{figure}[htb]
	\centering
	\begin{minipage}[b]{0.33\textwidth}
		\includegraphics[width=\textwidth]{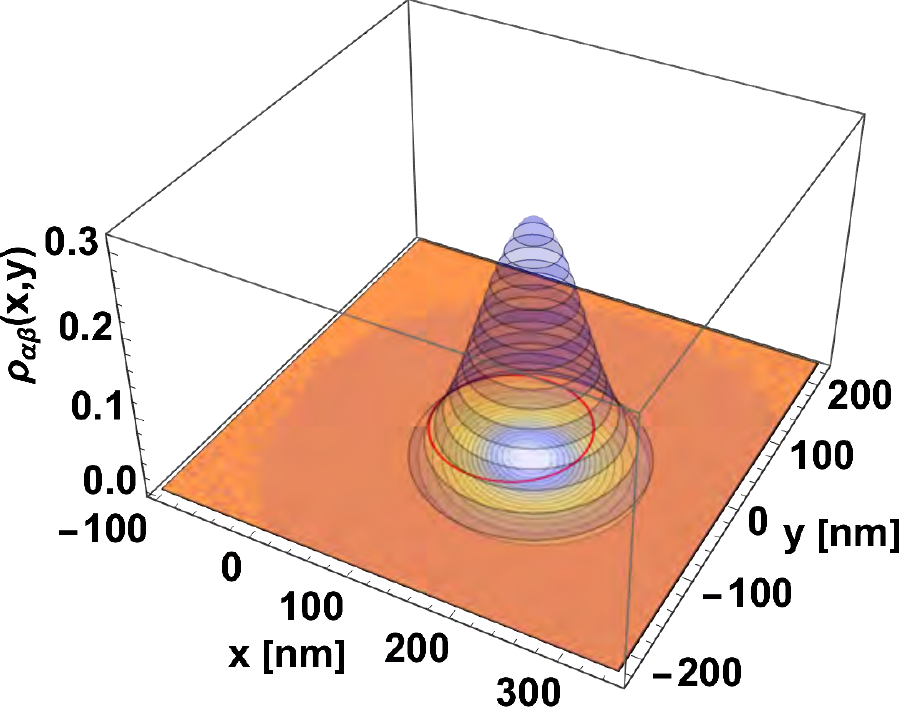}\\
		\centering{\footnotesize (a) $\alpha=\exp(i\pi/2)$, $\beta=2$.}
		\label{fig:rho_alphabeta1_a}
	\end{minipage}
\hspace{3cm}
	~ 
	\begin{minipage}[b]{0.33\textwidth}
		\includegraphics[width=\textwidth]{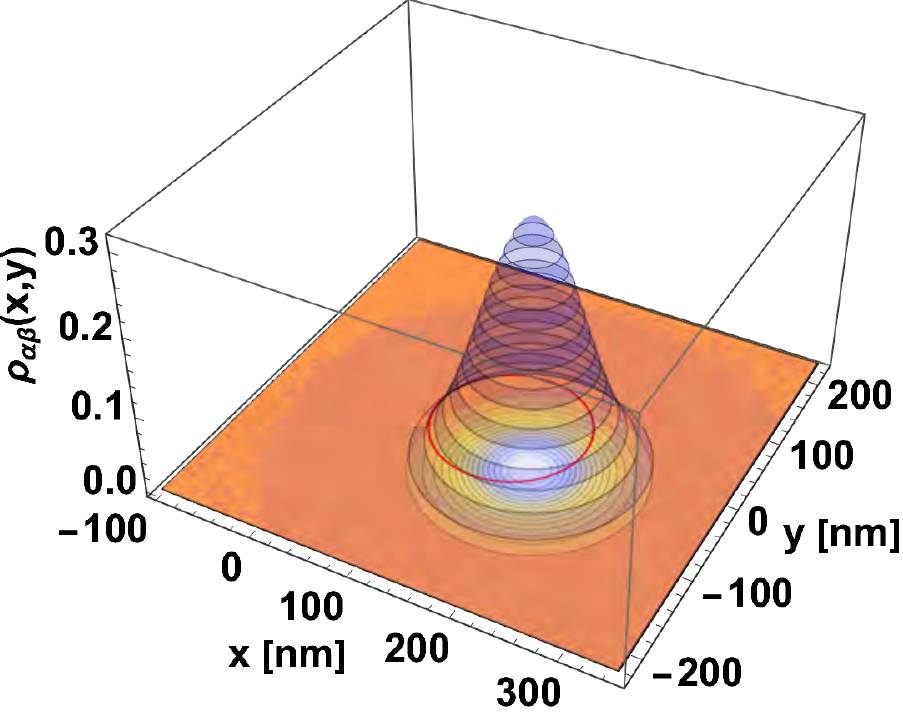}\\
		\centering{\footnotesize (b) $\alpha=\exp(i\pi/2)$, $\beta=2$.}
		\label{fig:rho_alphabeta1_b}
	\end{minipage}
	\\ [2ex]
	~ 
	\begin{minipage}[b]{0.33\textwidth}
		\includegraphics[width=\textwidth]{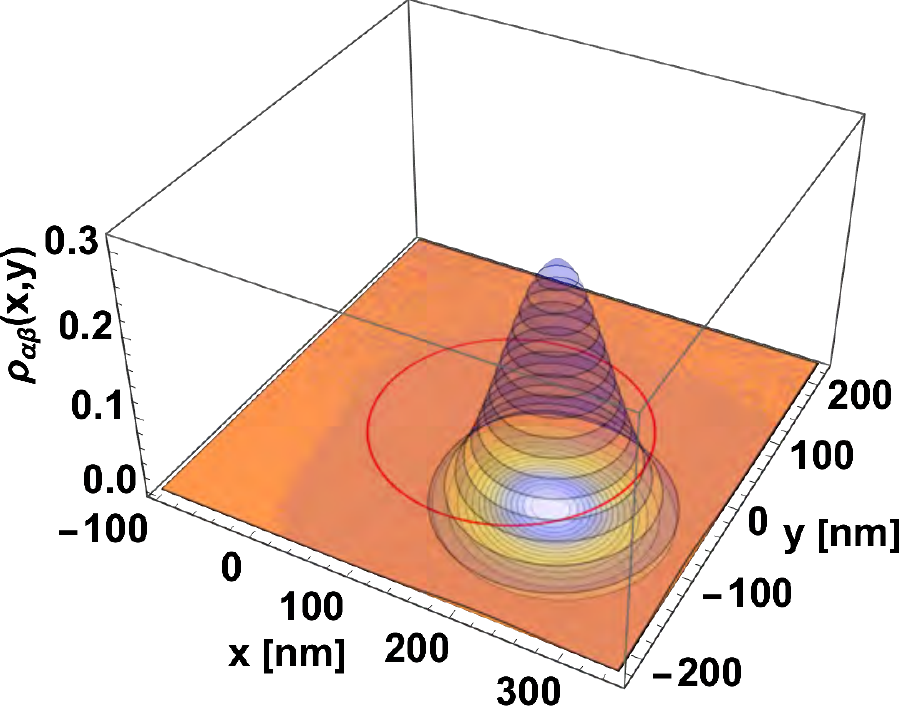}\\
		\centering{\footnotesize (c) $\alpha=2\exp(i\pi/2)$, $\beta=2$.}
		\label{fig:jtheta_alphabeta1_e}
	\end{minipage}
\hspace{3cm}
	~ 
	\begin{minipage}[b]{0.33\textwidth}
		\includegraphics[width=\textwidth]{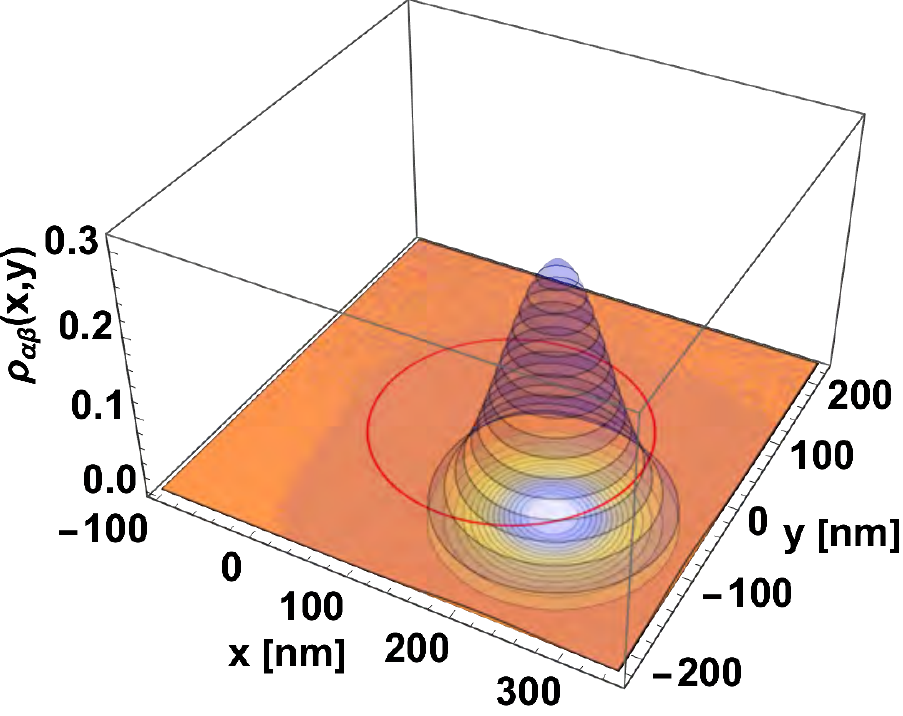}\\
		\centering{\footnotesize (d) $\alpha=2\exp(i\pi/2)$, $\beta=2$.}
		\label{fig:jtheta_alphabeta1_f}
	\end{minipage}
	\caption{\label{fig:rho_alphabeta1}Probability density $\rho_{\alpha,\beta}$ (3D plots), radial current density $j_{\alpha,\beta,\vec{u}_{\xi}}$ (2D plots in (a) and (c)), and angular current density $j_{\alpha,\beta,\vec{u}_{\theta}}/(e\,v_{\rm F})$ (2D plots in (b) and (d)) with $B_{0}=0.3$ T are shown for some 2D coherent states $\Upsilon_{\alpha,\beta}(x,y)$ in (\ref{84a}) setting $\delta=\pi/4$. The red line shows the classical trajectory (\ref{trajectory}) that the maximum of $\rho_{\alpha,\beta}$ would follow around the point $(x_{0},y_{0})$.}
\end{figure}

Finally, according to Eq.~(\ref{44}), a set of two-dimensional coherent states can be obtained through the correct composition of partial coherent states as follows \cite{d17,dosr20}:
\begin{equation}\label{83}
\Upsilon_{\alpha,\beta}=\mathcal{N}_{\alpha,\beta} \sum_{m=0}^{\infty}d_m^\beta\, \Pi_{m,\alpha}=\mathcal{N}_{\alpha,\beta} \sum_{n=0}^{\infty}c_n^\alpha\, \Pi_{\beta,n},
\end{equation}where $\mathcal{N}_{\alpha,\beta}$ are normalization constants and $\Pi_{\beta,n}$ and $\Pi_{m,\alpha}$ are the partial coherent states of the previous subsections.
Hence, employing the coherent states in (\ref{64}) and (\ref{78}), we obtain the corresponding two-dimensional coherent states,
\begin{equation}\label{84a}
\Upsilon_{\alpha,\beta}(x,y)=\frac{\exp\left(\left(\beta-\frac{z}{2}\right)z^\ast-\frac{\vert\beta\vert^2}{2}\right)}{\sqrt{\pi(2\exp(\vert\tilde{\alpha}\vert^2)-1)}}\sum_{n=0}^{\infty}\frac{\tilde{\alpha}^n}{n!}\left(\begin{array}{c}
	\sqrt{n}(z-\beta)^{n-1}\\
	i(z-\beta)^{n}
	\end{array}\right),\quad z= \frac{\sqrt{2}}{\ell_{\rm B}} \left(\frac{x+iy}{2}\right),\quad \tilde{\alpha}=\alpha e^{-i\delta},
\end{equation}
as well as their corresponding probability and current densities, which are illustrated in Figure \ref{fig:rho_alphabeta1}:
\begin{eqnarray}
 	\rho_{\alpha,\beta}(x,y)\!&\!=\!&\! \frac{\exp\left(-\vert z-\beta\vert^2\right)}{\pi(2\exp(\vert\tilde{\alpha}\vert^2)-1)}\left[1+\left\vert\sum_{n=1}^{\infty}\frac{\left(\tilde{\alpha}(z-\beta)\right)^n}{n!}\right\vert^2+\left\vert\sum_{n=1}^{\infty}\frac{\left(\tilde{\alpha}(z-\beta)\right)^n\sqrt{n}}{n!(z-\beta)}\right\vert^2
 +2\, \textrm{Re}\!\left(\sum_{n=1}^{\infty}\frac{\left(\tilde{\alpha}(z-\beta)\right)^n}{n!}\right) \right]\!, \quad 
 \label{84b} 
 \\ [2ex]
 	j_{\alpha,\beta,\vec{u}}(x,y)\!&\!=\!&\! \frac{2ev_{\rm F}\,\exp\left(-\vert z-\beta\vert^2\right)}{\pi(2\exp(\vert\tilde{\alpha}\vert^2)-1)}\, \textrm{Re}\Bigg( i(-i)^{k}e^{-i\theta}\left(\sum_{n'=0}^{\infty}\frac{\left[\tilde{\alpha}(z-\beta)\right]^{n'}}{n'!}\right)  \left(\sum_{n=0}^{\infty}\frac{\left(\tilde{\alpha}^\ast(z^\ast-\beta^\ast)\right)^n\sqrt{n}}{n!(z^\ast-\beta^\ast)}\right)\Bigg).
	\label{84c}
	\end{eqnarray}

The mean energy $\langle H_{\rm DW}\rangle_{\alpha,\beta}$ now has an identical behavior as that in Eq.~(\ref{82}) because the contribution of the partial coherent states $\Pi_{\beta,n}(x,y)$ is the same as that of $\Psi_{m,n}(x,y)$.

\subsubsection{Cyclotron motion}

On the other hand, the mean values of the operators in Eq.~(\ref{matrixoperators}) for the two-dimensional coherent states $\Upsilon_{\alpha,\beta}$ are
\begin{eqnarray}
	\langle\mathcal{U}_{q}\rangle_{\alpha,\beta}&=\frac{\beta+(-1)^{q}\beta^{\ast}}{\sqrt{2}i^{q}}, \qquad \langle\mathcal{V}_{q}\rangle_{\alpha,\beta}&=\frac{i^{q}(\tilde{\alpha}+(-1)^{q}\tilde{\alpha}^{\ast})}{\sqrt{2}(2\exp\left(\vert\tilde{\alpha}\vert^2\right)-1)}\left(\exp\left(\vert\tilde{\alpha}\vert^{2}\right)+\sum_{n=1}^{\infty}\frac{\vert\tilde{\alpha}\vert^{2n}}{\sqrt{(n-1)!(n+1)!}}\right), \\ [1ex]
	\langle\mathcal{R}_{0}^{2}\rangle_{\alpha,\beta}&=2\vert\beta\vert^{2}+1, \quad \langle(\mathcal{R}')^{2} \rangle_{\alpha,\beta}&=\frac{4\vert\tilde{\alpha}\vert^{2}\exp\left(\vert\tilde{\alpha}\vert^{2}\right)+1}{2\exp\left(\vert\tilde{\alpha}\vert^2\right)-1}.
\end{eqnarray}
Here, the above mean values coincide with those for the partial coherent states $\Pi_{\beta,n}$ and $\Pi_{m,\alpha}$ when one takes the sums over the indices $n$ and $m$, respectively.

As we can see in Figure~\ref{fig:rho_alphabeta1}, the complex parameters $\alpha$ and $\beta$ determine again where the maximum probability amplitude of the coherent states is. The kind of two-dimensional coherent states given in Eq.~(\ref{84a}) exhibits a stable Gaussian probability distribution independently on the value of $\alpha$, so that they resemble the standard  harmonic oscillator coherent states represented in phase space. Regarding a physical interpretation, the description given in \cite{fk70} is valid, in general terms, for the case discussed here: while $\beta$ determines the position $(x_{0},y_{0})$ respect to the origin of the classical trajectory center, $\alpha$ indicates the position of the Gaussian package around the point $(x_0,y_0)$:
\begin{equation}\label{trajectory}
	\left(x-\sqrt{2}\ell_{\rm B}\, \textrm{Re}(\beta)\right)^{2}+\left(y-\sqrt{2}\ell_{\rm B}\, \textrm{Im}(\beta)\right)^{2}=\frac{\ell_{\rm B}^{2}\left(4\vert\tilde{\alpha}\vert^{2}\exp\left(\vert\tilde{\alpha}\vert^{2}\right)+1\right)}{2\exp\left(\vert\tilde{\alpha}\vert^2\right)-1}.
\end{equation}

\begin{figure}[htb]
	\centering
	\includegraphics[width=.34\linewidth]{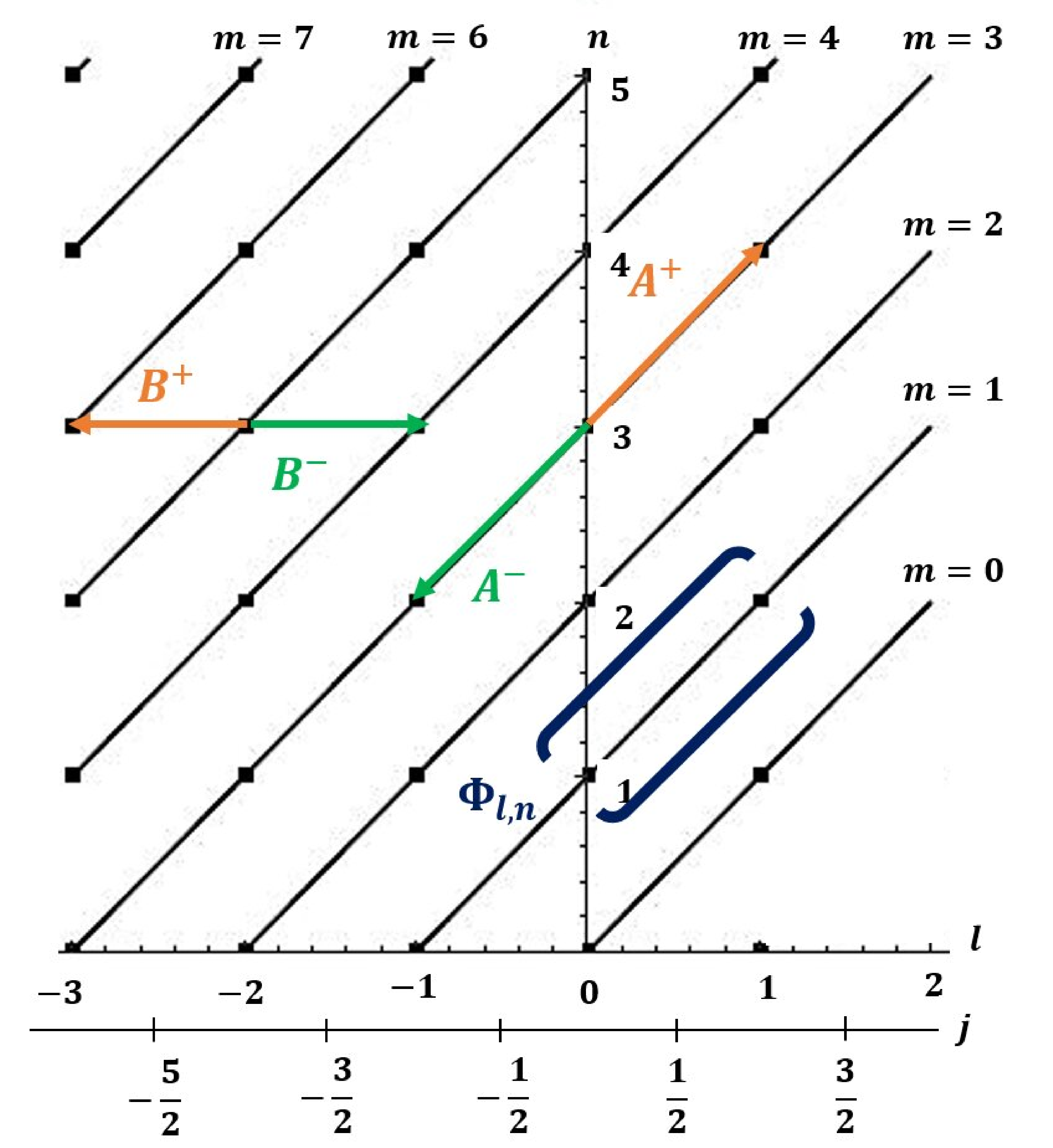}
	\caption{\label{fig:diagramLN} 
		The space of scalar states $\phi_{l,n}$, where the point $(l,n)$ identifies the state $\phi_{l,n}$. Slanted lines connect states with the same value of $m=n-l$.}
\end{figure}

\section{Coherent states with a fixed ``total angular momentum"}\label{sec4}

Each pseudo-spinor eigenstate $\Psi_{m,n}$, as given in (\ref{32}) and (\ref{33}), has components $\psi_{m,n-1}$ and $\psi_{m,n}$, each with angular momentum eigenvalues $l-1=n-m-1$ and $l=n-m$, respectively. Although the angular momentum of $\Psi_{m,n}$ is not well defined, its ``total angular momentum" $\mathbb{J}_{z}$ as defined in (\ref{36}) has eigenvalue $j=l-1/2$, which is the half sum of the $L_{z}$ values of the two components. Therefore, to make explicit the value of the total angular momentum of $\Psi_{m,n}$ and the orbital momentum of its components,  in this section we will use the following notation for the eigenstates:
\begin{equation}
 \Phi_{l,n}\equiv\Psi_{m,n}, \quad \phi_{l,n}\equiv\psi_{m,n}, \quad \phi_{l-1,n-1}\equiv\psi_{m,n-1}, \quad \textrm{with} \quad l=n-m\leq n,
\end{equation}
with
\begin{equation}\label{88}
\Phi_{l,n}(x,y)=\frac{1}{\sqrt{2^{1-\delta_{0n}}}}\left(\begin{array}{c}
(1-\delta_{0n})\phi_{l-1,n-1}(x,y) \\
i\,\phi_{l,n}(x,y)
\end{array}\right).
\end{equation}
A scheme of the new notation can be seen in Fig.~\ref{fig:diagramLN}, that may be compared with Fig.~\ref{fig:diagram}.
By $\Phi_{l,n}$ we denote an state with total angular momentum $j=l-1/2$ and energy $\hbar\,\omega\sqrt{n}$. From (\ref{31}), the explicit form of the component $\phi_{l,n}$, $n=0,1,\dots$, $l=0\pm1,\dots$, is
\begin{equation}\label{89}
  \phi_{l,n}(\xi,\theta)=\frac1{\ell_{\rm B}}(-1)^{\min(n-l,n)}\sqrt{\frac{1}{2\pi}\frac{\min(n-l,n)!}{\max(n-l,n)!}}\,  \xi^{\vert l\vert}
\, e^{-\frac{\xi^2}{2}+il\theta}\,   L_{\min(n-l,n)}^{\vert l\vert}\left(\xi^2\right).
\end{equation}

If we fix the value of $l\equiv n-m$ of the pseudo-spinor states $\Phi_{l,n}$, one can construct pseudo-spinor coherent states $\Xi_{j,\zeta}$ that satisfy the eigenvalue equation~(\ref{36}) by means of linear combinations of pseudo-spinor states $\Phi_{l,n}(x,y)$ with different values of $n$ and the same $l=j+1/2$. For that purpose, let us consider the following operators
	\begin{equation}\label{86}
 \mathbb{K}^-=\mathbb{A}^-\mathbb{B}^-=\left[\begin{array}{c c}
\cos\delta\, \frac{\sqrt{N+2}}{\sqrt{N+1}}A^-B^- & \sin\delta\, \frac{1}{\sqrt{N+1}}(A^-)^2B^- \\ [1.5ex]
-\sin\delta\, \sqrt{N+1}B^- & \cos\delta\,  A^-B^-
\end{array}\right], \quad \mathbb{K}^+=(\mathbb{K}^-)^\dagger.
\end{equation}
They satisfy
\begin{equation}
 [\mathbb{K}^-,\mathbb{K}^+]\equiv2\mathbb{K}_0=\left[\begin{array}{c c}
N+M+2 & 0 \\
0 & N+M+1
\end{array}\right], \quad [\mathbb{K}_0,\mathbb{K}^\pm]=\pm\mathbb{K}^\pm, \label{87a} 
\end{equation}
which allow us to identify the {\it su$(1,1)$} algebra generated by the operators $\mathbb{K}^\pm$, $\mathbb{K}_0$. 
We also have,
\begin{equation}
 \mathbb{K}^-\Phi_{l,n}=\frac{1}{\sqrt{2^{\delta_{1n}}}}\sqrt{n(n-l)}\,  e^{i\delta}\, \Phi_{l,n-1}, \quad n=0,1,2,\dots, \quad l=0,\pm 1,\pm 2,\dots \label{87b}
\end{equation}
In a similar way as in the previous section, for the special case $n=0$ we must define in a proper way the operators $\mathbb{K}^{\pm}$. Thus, we can obtain excited pseudo-spinor states with fixed $j$ by applying the creation operators on two types of ground states, corresponding to $j>0$ or $j<0$, as follows:
\begin{eqnarray}
(\mathbb{K}^+)^{k}\Phi_{l,0}^- =\frac{\sqrt{2(k-l)!k!}}{\sqrt{(-l)!}}\exp\left(-ik\delta\right)\Phi_{l,k}^-, \quad l\leq0, 
\qquad
(\mathbb{K}^+)^k\Phi_{l,l}^+ =\frac{\sqrt{(k+l)!k!}}{\sqrt{l!}}\exp\left(-ik\delta\right)\,\Phi_{l,k+l}^+, \quad l>0. \label{92}
\end{eqnarray}

The pseudo-spinor coherent states $\Xi_{j,\zeta}$ are built as the common eigenstates of the annihilation operator $\mathbb{K}^-$ and the total angular momentum operator $\mathbb{J}_z$, {\it i.e.},
\begin{equation}
\mathbb{K}^-  \Xi_{j,\zeta} =\zeta\, \Xi_{j,\zeta}, \quad  \zeta\in\mathbb{C}, \qquad\qquad 
	\mathbb{J}_z\, \Xi_{j,\zeta} =j\, \Xi_{j,\zeta}, \quad  j=l-1/2. \label{94b}
\end{equation}
It is important to remark that the coherent states thus constructed resemble the so-called ``charged coherent states'' \cite{bbdr76}, where the scalar operator $L_z=A^+A^--B^+B^-$ is interpreted as the charge operator \cite{f04,ahb15}. Remark that there are two kinds of coherent states depending on the type of ground state. We can relate these eigenvalues to the classical motion of the charged particles. Since the classical motion is a circle, electrons move counterclockwise around the direction of the magnetic field $\vec{B}$. This means that the classical motion corresponds to $j\geq1/2$. We will focus on this case in the next section and for completeness we will briefly mention the case with $j<0$ at the end.

\begin{figure}[htb]
	\centering
	\begin{minipage}[b]{0.33\textwidth}
		\includegraphics[width=\textwidth]{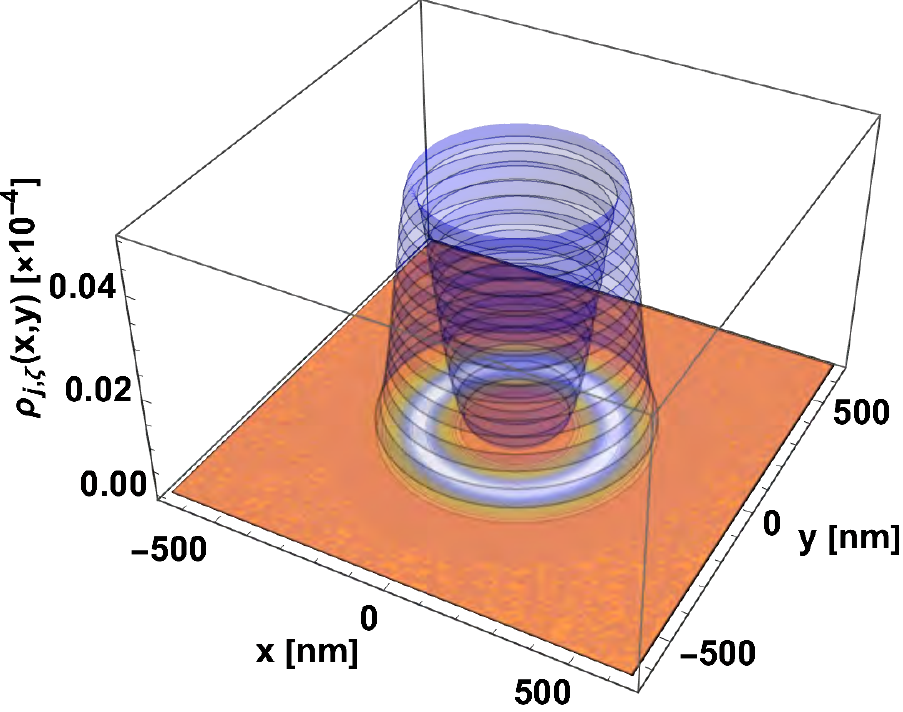}\\
		\centering{\footnotesize (a) $j=-5/2$}
		\label{fig:jtheta1P_zl_1}
	\end{minipage}
	\hspace{3cm}
	~ 
	\begin{minipage}[b]{0.33\textwidth}
		\includegraphics[width=\textwidth]{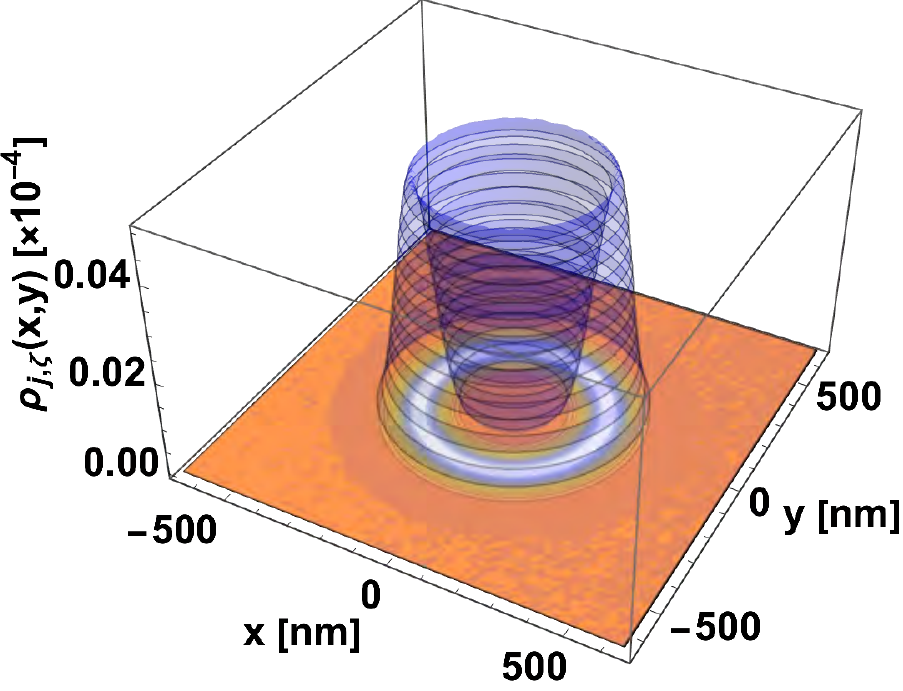}\\
		\centering{\footnotesize (b) $j=5/2$}
		\label{fig:jtheta1P_zl_2}
	\end{minipage}
	\\ [2ex]
	~ 
	\begin{minipage}[b]{0.33\textwidth}
		\includegraphics[width=\textwidth]{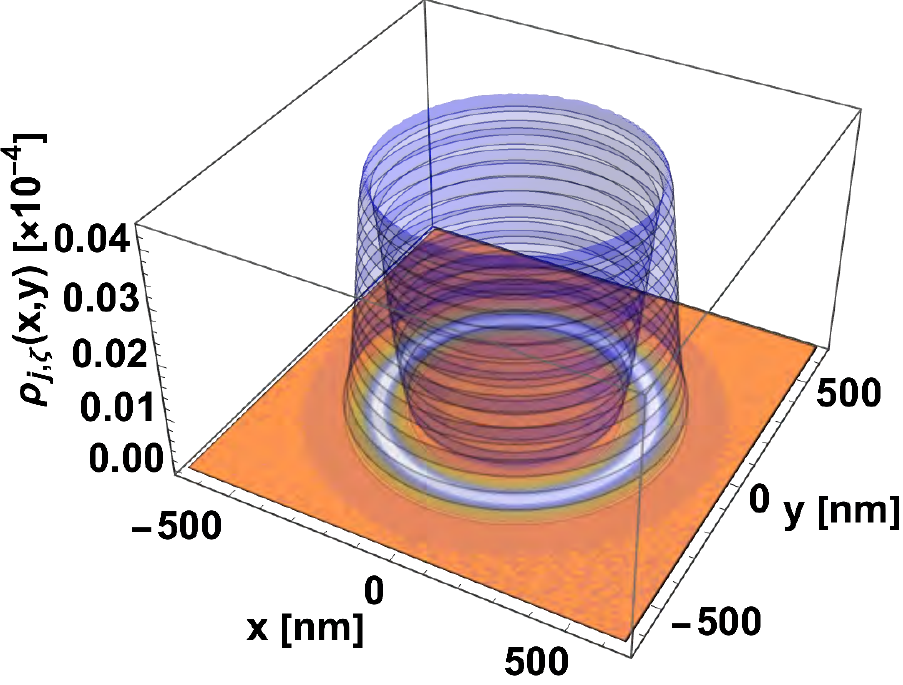}
		\\
		\centering{\footnotesize (c) $j=51/2$}
		\label{fig:jtheta1P_zl_3}
	\end{minipage}
	\hspace{3cm}
	~ 
	\begin{minipage}[b]{0.33\textwidth}
		\includegraphics[width=\textwidth]{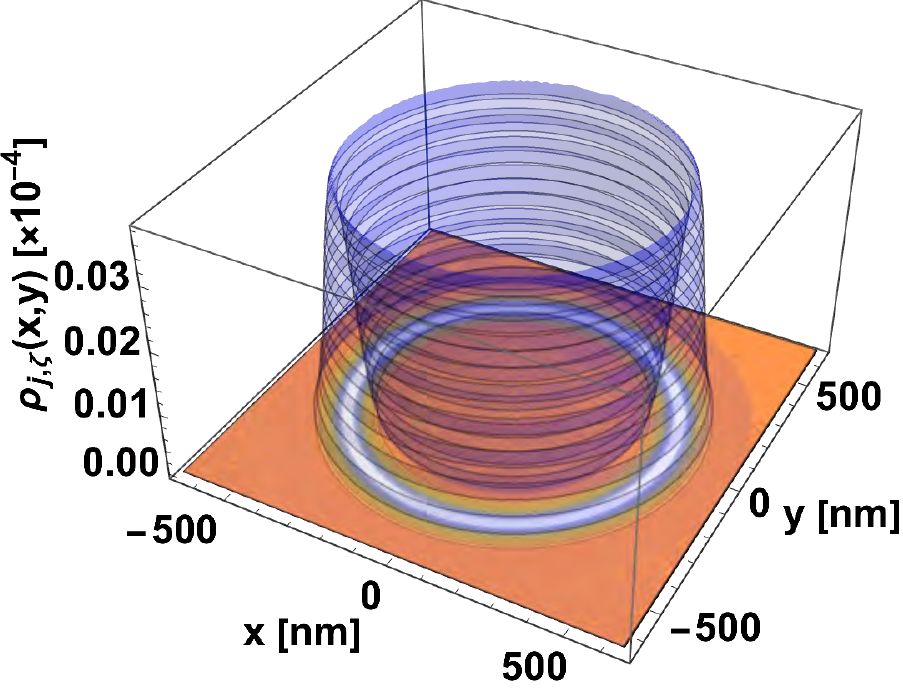}\\
		\centering{\footnotesize (d) $j=81/2$}
		\label{fig:H1P_z}
	\end{minipage}
	\caption{\label{fig:rho1P_z} Plots of the probability density $\rho_{j,\zeta}(x,y)$ (3D plots) and the 
		angular density $j_{j,\zeta,\vec{u}_\theta}/(e\,v_{\rm F})$ (2D plots) are shown for the coherent states $\Xi_{j,\zeta}$ in \eqref{97} and \eqref{95} with $\delta=\pi/4$ and some values of $j$. In all the cases $\zeta=5\exp(i\pi/2)$ and $B_{0}=0.3$ T.}
\end{figure}

\subsection{Coherent states with $j>0$}\label{sec4.1}

\begin{figure}[htb]
	\centering
	\begin{minipage}[b]{0.33\textwidth}
		\includegraphics[width=\textwidth]{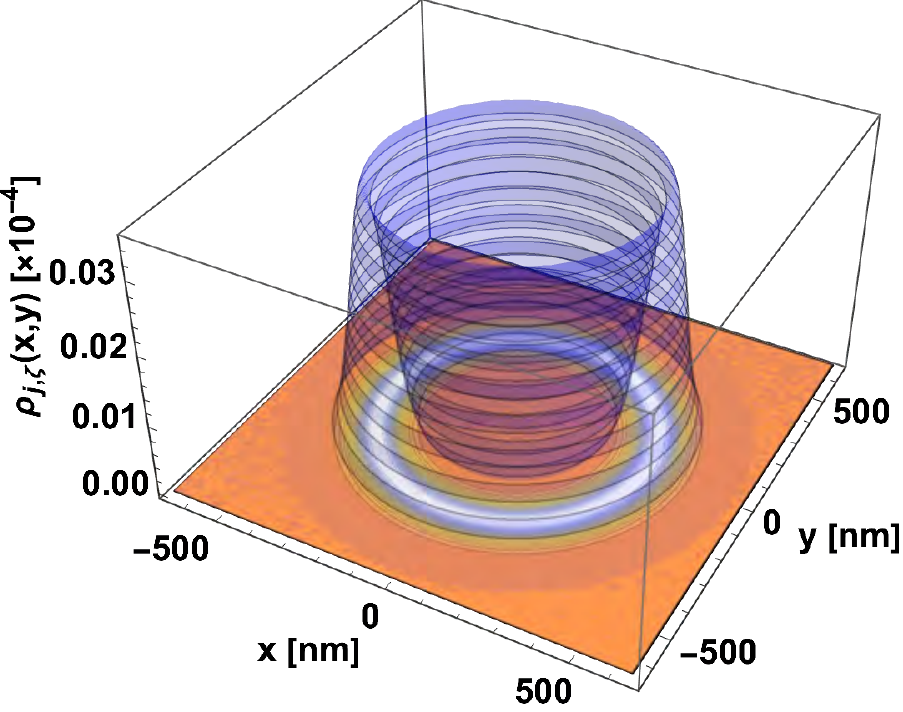}\\
		\centering{\footnotesize (a) $j=-5/2$}
		\label{fig:jtheta1N_zl_1}
	\end{minipage}
\hspace{3cm}
	~ 
	\begin{minipage}[b]{0.33\textwidth}
		\includegraphics[width=\textwidth]{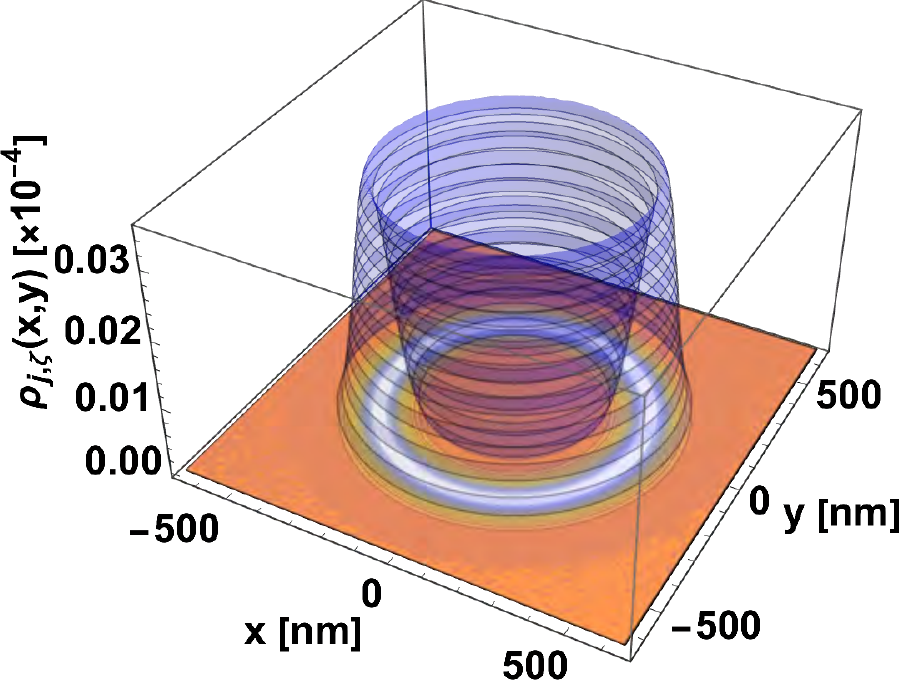}\\
		\centering{\footnotesize (b) $j=5/2$}
		\label{fig:jtheta1N_zl_2}
	\end{minipage}
	\\  [2ex]
	~ 
	\begin{minipage}[b]{0.33\textwidth}
		\includegraphics[width=\textwidth]{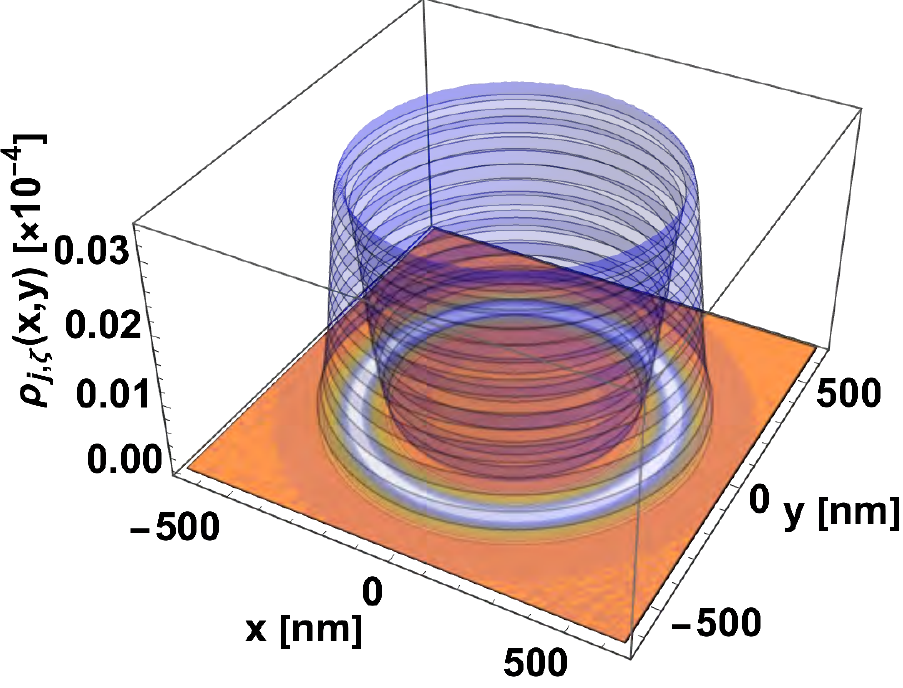}\\
		\centering{\footnotesize (c) $j=51/2$}
		\label{fig:jtheta1N_zl_3}
	\end{minipage}
\hspace{3cm}
	~ 
	\begin{minipage}[b]{0.33\textwidth}
		\includegraphics[width=\textwidth]{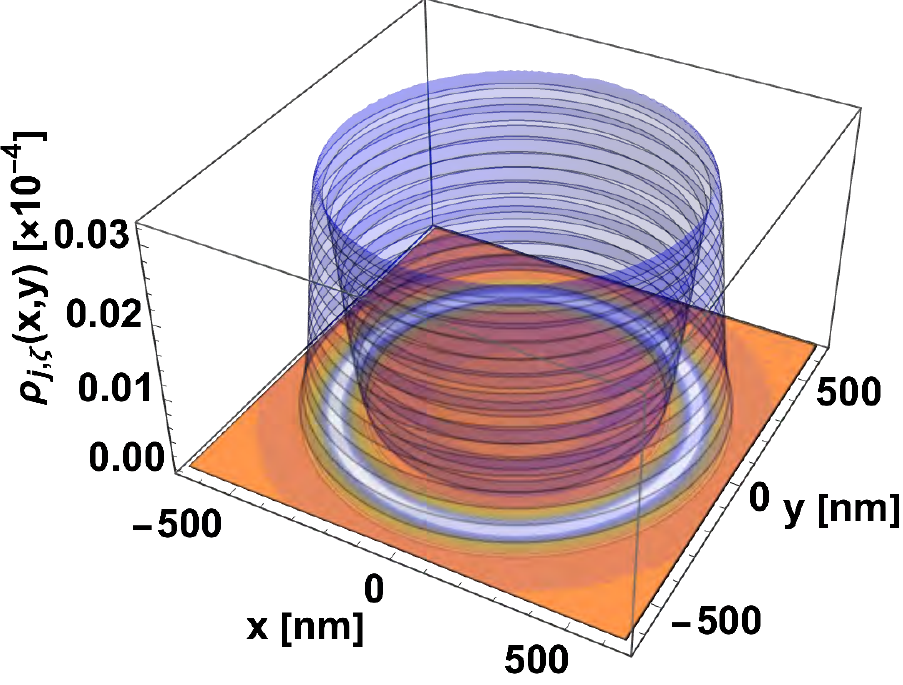}\\
		\centering{\footnotesize (d) $j=81/2$}
		\label{fig:H1N_z}
	\end{minipage}
	\caption{\label{fig:rho1N_z}Plots of the probability density $\rho_{j,\zeta}(x,y)$ (3D plots) and the angular density 
	$j_{j,\zeta,\vec{u}_\theta}/(e\,v_{\rm F})$ (2D plots) are shown for the coherent states $\Xi_{j,\zeta}$ in \eqref{97} and \eqref{95} with $\delta=\pi/4$ and some values of $j$. In all the cases $\zeta=10\exp(i\pi/2)$ and $B_{0}=0.3$ T.}
\end{figure}

\begin{figure}[htb]
	\centering
	\begin{minipage}[b]{0.34\textwidth}
		\includegraphics[width=\textwidth]{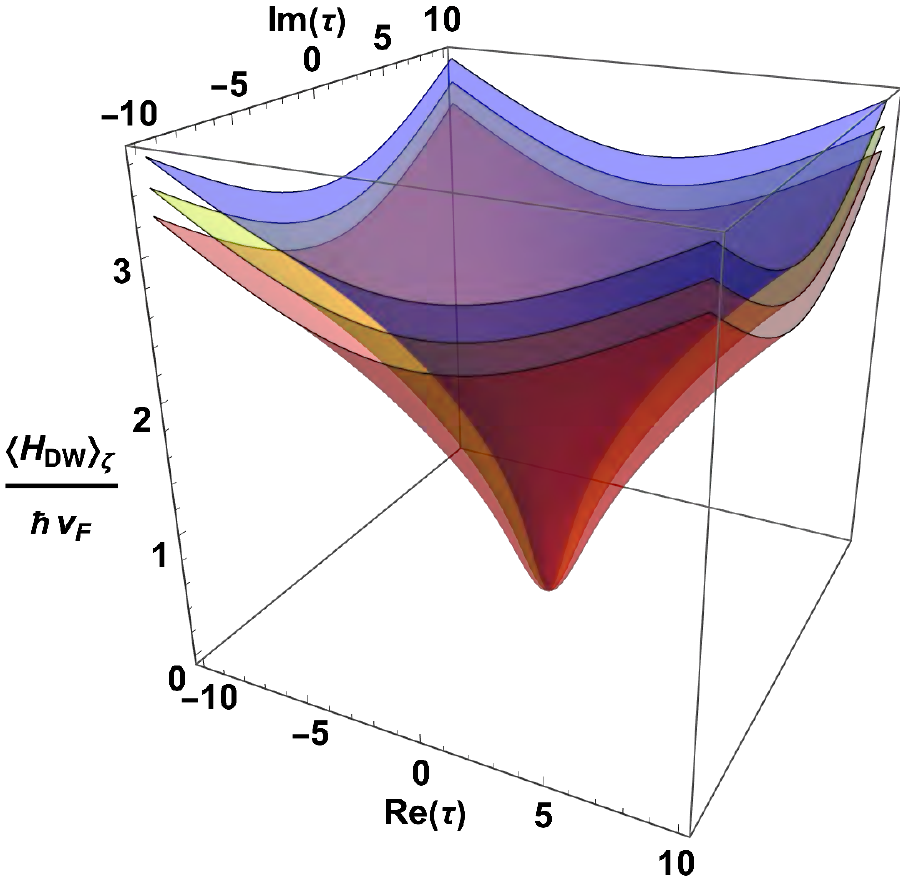}
		\label{fig:H1P_z}
	\end{minipage}
\hspace{3cm}
	\begin{minipage}[b]{0.34\textwidth}
		\includegraphics[width=\textwidth]{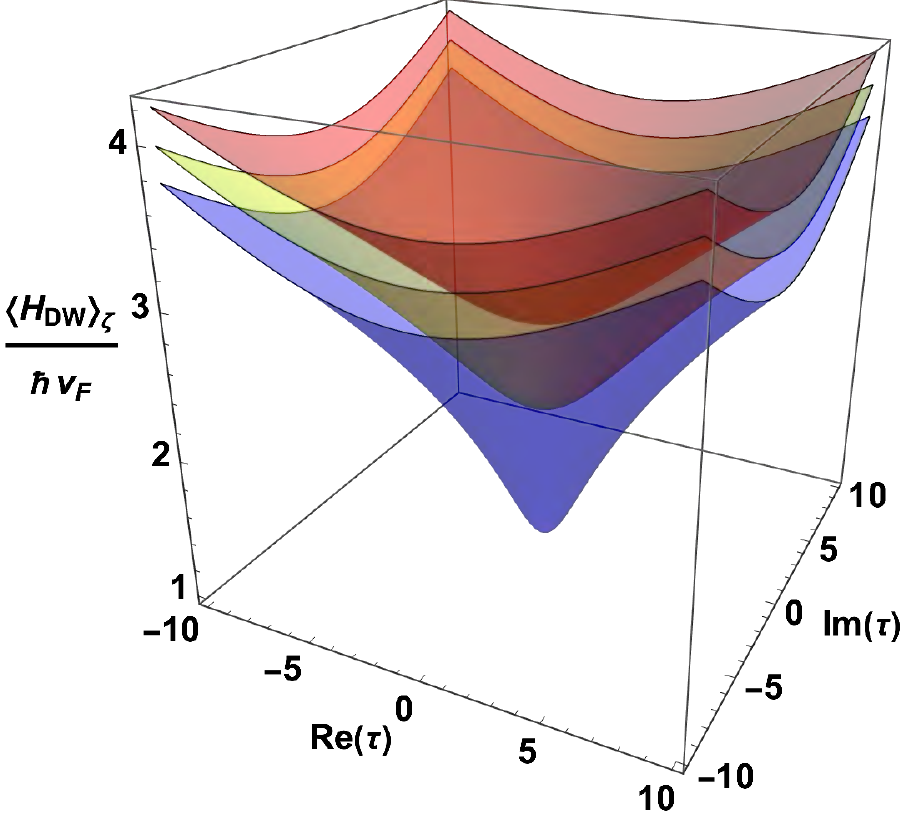}
		\label{fig:H1N_z}
	\end{minipage}
	\caption{\label{fig:H_z}Mean energy value $\langle H_{\rm DW}\rangle_\zeta/(\hbar v_{\rm F})$ with 
	$B_{0}=0.3$ T as a continuous function of $\zeta$ for the coherent states $\Xi_{j,\zeta}$ is shown for 
	some values of $l$: $\vert l\vert=1$ (blue), $\vert l\vert=4$ (yellow), $\vert l\vert=7$ (red).
	On the left $j<0$ and on the right $j>0$.}
\end{figure}

For the first quadrant in Fig.~\ref{fig:diagramLN} ($l>0, n\geq 0$), the ground states are $\Phi_{l,l}$ and the pseudo-spinor coherent states are
\begin{equation}\label{97}
\Xi_{j,\zeta}(x,y) =\left({}_0F_1\left(l+1;\vert {\zeta}\vert^2\right)\right)^{-1/2} \sum_{n=l}^{\infty}\frac{\sqrt{l!}\ \tilde{\zeta}^{n-l}}{\sqrt{n!(n-l)!}}\, \Phi_{l,n} (x,y), \qquad \tilde{\zeta}=\zeta\, e^{-i\delta},
\end{equation}
where ${}_0F_1$ denotes the confluent hypergeometric function. The corresponding probability and current densities, as well as the mean energy value, are given by
\begin{eqnarray}
\hskip-0.8cm 	&&
\rho_{j,\zeta}(\xi,\theta)=\Xi^{\dagger}_{j,\zeta} \Xi_{j,\zeta}= \frac{\vert z\vert^{2l}\exp\left(-\xi^2\right)}{4\pi\ell_{\rm B}^{2}\, _0F_1\left(l+1;\vert\tilde{\zeta}\vert^2\right)}
 \left[ \left\vert\sum_{n=l}^{\infty}\frac{\sqrt{l!}(-\tilde{\zeta})^{n-l}}{n!}L_{n-l}^{l}(\xi^2)\right\vert^2 +\left\vert\sum_{n=l}^{\infty}\frac{\sqrt{l!}(-\tilde{\zeta})^{n-l}z^{-1}}{\sqrt{n}(n-1)!}L_{n-l}^{l-1}(\xi^2)\right\vert^2\right],  
 \label{98a} 
  \\ [1ex]
\hskip-0.8cm 	&&
j_{j,\zeta,\vec{u}}(\xi)=ev_{\rm F}\, \Xi^{\dagger}_{j,\zeta}\,(\vec{\sigma}\cdot\vec{u})_k\,
\Xi_{j,\zeta}=\frac{ev_{\rm F}\,\vert z\vert^{2l}\exp\left(-\xi^2\right)}{2\pi\ell_{\rm B}^{2} \, _0F_1\left(l+1;\vert\tilde{\zeta}\vert^2\right)} \nonumber \\
\hskip-0.8cm 	&\, & \hskip4.5cm \times\, \textrm{Re}\left[ i(z^\ast)^{-1}(-i)^{k}e^{-i\theta}
\sum_{n'=l}^{\infty}\frac{\sqrt{l!}(-\tilde{\zeta})^{n'-l}}{n'!}L_{n'-l}^{l}(\xi^2)
\sum_{n=l}^{\infty}\frac{\sqrt{l!}(-\tilde{\zeta}^\ast)^{n-l}}{\sqrt{n}(n-1)!}L_{n-l}^{l-1}(\xi^2)
\right],\label{98b}
  \\ [1ex]
\hskip-0.8cm 	&&
\langle H_{\rm DW}\rangle_\zeta=\frac{\hbar\,\omega}{\, _0F_1\left(l+1;\vert\tilde{\zeta}\vert^2\right)}\sum_{n=l}^{\infty}\frac{l!\,\vert\tilde{\zeta}\vert^{2n-2l}}{n!(n-l)!}\sqrt{n}. \label{98c}
\end{eqnarray}
In Figures \ref{fig:rho1P_z} and  \ref{fig:rho1N_z}, plots of the probability density $\rho_{j,\zeta}$ and the angular density 
$j_{j,\zeta,\vec{u}_\theta}$ are shown. As we can see, the probability density is basically a ring centered at the origin whose radius increases as $j$ grows. More precisely, the values of both $\zeta$ and $j$ modify the probability density shape as well the angular current density behavior: the maximum values of both functions move away radially from the origin as the parameters $\vert\zeta\vert$ and $\vert j\vert$ increase.
In Figure~\ref{fig:H_z} a plot of the mean value of the energy in a coherent state $\Xi_{{j,\zeta}}(x,y)$ is given.

\subsubsection{Cyclotron motion}

	Finally, the mean values of the operators in Eq.~(\ref{matrixoperators}) for the coherent states $\Xi_{j,\zeta}$ are given by:
	\begin{eqnarray}
		 \langle\mathcal{U}_{q}\rangle_{j,\zeta}=0,\hskip3.87cm & &\qquad\; \langle\mathcal{V}_{q}\rangle_{j,\zeta}=\frac{i^{q}(\tilde{\zeta}+(-1)^{q}\tilde{\zeta}^{\ast})}{\sqrt{2}\,_0F_1\left(l+1;\vert\tilde{\zeta}\vert^2\right)}\sum_{n=l}^{\infty}\frac{l!\sqrt{n-l+1}\vert\tilde{\zeta}\vert^{2(n-l)}}{n!(n-l+1)!}, \\
		\langle\mathcal{R}_{0}^{2}\rangle_{j,\zeta}=\frac{2\ l!\ I_{l+1}(2\vert\tilde{\zeta}\vert)}{\vert\tilde{\zeta}\vert^{l-1}\,_0F_1\left(l+1;\vert\tilde{\zeta}\vert^2\right)}+1, &&\quad 
		\langle(\mathcal{R}')^{2} \rangle_{j,\zeta}=2l\ \frac{_0F_1\left(l;\vert\tilde{\zeta}\vert^2\right)}{\,_0F_1\left(l+1;\vert\tilde{\zeta}\vert^2\right)},
	\end{eqnarray}
where $I_{k}(z)$ denotes the modified Bessel function of the first kind. In comparison with the coherent states built above, the classical position of electrons is also determined by the real and imaginary parts of the corresponding eigenvalue ($\zeta$, in this case), while the mean value of the operator for the classical circular trajectory $\langle(\mathcal{R}')^{2} \rangle_{j,\zeta}$ depends explicitly on the positive $z$-component of the angular momentum $l$, which agrees with the behavior of the probability density $\rho_{j,\zeta}$ shown in Fig.~\ref{fig:rho1P_z}.

\subsection{Coherent states with $j<0$}
\label{sec4.2}
Now, for the second quadrant ($l\leq 0, n\geq 0$) in Figure~\ref{fig:diagramLN}, where the ground states are $\Phi_{l,0}$, we have
\begin{equation}\label{95}
 	\Xi_{j,\zeta}(x,y)= \frac1{\sqrt{2 \, _0F_1\left(-l+1;\vert \tilde{\zeta}\vert^2\right)-1}}\, \left(\Phi_{l,0}(x,y)+\sum_{n=1}^{\infty}\frac{\sqrt{2\,(-l)!}\tilde{\zeta}^n}{\sqrt{n!(n-l)!}}\ \Phi_{l,n}(x,y)\right).
\end{equation}
The corresponding probability and current densities, as well the mean energy value are given by
\begin{eqnarray}
 	 \rho_{j,\zeta}(\xi,\theta)&=& \Xi^{\dagger}_{j,\zeta} \Xi_{j,\zeta}	 =\frac{\vert z\vert^{-2l}\exp\left(-\xi^2\right)/(2\pi\ell_{\rm B}^{2})}{ 2 \, _0F_1\left(-l+1;\vert\tilde{\zeta}\vert^2\right)-1 }\Bigg\{ 2\, \textrm{Re}\left( \sum_{n=1}^{\infty}\frac{(-\tilde{\zeta})^n}{(n-l)!}L_{n}^{-l}(\xi^2)\right) \label{96a} 
	 \\
 	&&  \qquad\qquad\qquad\qquad + \frac{1}{(-l)!}+\left\vert\sum_{n=1}^{\infty}\frac{\sqrt{(-l)!}(-\tilde{\zeta})^n}{(n-l)!}L_{n}^{-l}(\xi^2)\right\vert^2+\left\vert\sum_{n=1}^{\infty}\frac{\sqrt{(-l)!}(-\tilde{\zeta})^n}{(n-l)!}\frac{z^\ast}{\sqrt{n}}L_{n-1}^{-l+1}(\xi^2)\right\vert^2\Bigg\}, \nonumber 
\\ [1ex]
	   j_{j,\zeta,\vec{u}}(\xi) &=& ev_{\rm F}\, \Xi^{\dagger}_{j,\zeta} \,(\vec{\sigma}\cdot\vec{u})_k\, \Xi_{j,\zeta} =-\frac{ev_{\rm F}\,\vert z\vert^{-2l}\exp\left(-\xi^2\right)}{\pi\ell_{\rm B}^{2}\left[2 \, _0F_1\left(-l+1;\vert\tilde{\zeta}\vert^2\right)-1\right]}  \label{96b}  \\
 	&&\qquad\qquad\qquad\qquad  \times\,\textrm{Re}\left( iz(-i)^{k}e^{-i\theta} \sum_{n'=0}^{\infty}\frac{\sqrt{(-l)!}(-\tilde{\zeta})^{n'}}{(n'-l)!}L_{n'}^{-l}(\xi^2) \sum_{n=1}^{\infty}\frac{\sqrt{(-l)!}(-\tilde{\zeta}^\ast)^{n}}{\sqrt{n}(n-l)!}L_{n-1}^{-l+1}(\xi^2)  \right), \nonumber
\\ [1ex]
 	\langle H_{\rm DW}\rangle_\zeta &=& \frac{2\hbar\,\omega}{2\,_0F_1\left(-l+1;\vert\tilde{\zeta}\vert^2\right)-1}\ \sum_{n=0}^{\infty}\frac{(-l)!\,\vert\tilde{\zeta}\vert^{2n}}{n!(n-l)!}\sqrt{n}. \label{96c}
\end{eqnarray}
It is important to remark that the values of the radial current density $j_{j,\zeta,\vec{u}_{\xi}}$ are negligible for all the coherent states with fixed total angular momentum, so that there is a very low probability of flux in the radial direction.
On the other hand, as the total angular momentum $\vert j\vert$ increases for coherent states with $j<0$, the corresponding mean energy value $\langle H_{\rm DW}\rangle_\zeta$ takes smaller values while for the states with $j>0$ the opposite effect occurs  (see Fig.~\ref{fig:H_z}). This seems reasonable according to the pseudo-spinor composition of the two types of coherent states.

\section{Conclusions}\label{sec5}
In this work, we have applied the Barut-Girardello formalism to construct the coherent states for the physical system that arises from the interaction between electrons in a graphene layer that lies on the $x-y$ plane and a constant magnetic field directed along the $z$-axis. Since we want to examine the semi-classical states with rotational symmetry, we have used a symmetric gauge of the potential, first to solve the physical problem in polar coordinates and to identify the relevant annihilation operators, and then to construct the coherent states as eigenstates of such operators.

This system has pseudo-spinor eigenstates 	$\Psi_{m,n}$ that are labeled by two positive integers: $n$ for the energy level while $m$ labels de infinite degeneracy of each level. Associated to these solutions there are two commuting sets of creation-annihilation operators, $A^{\pm}$ and $B^{\pm}$. Due to the two components of the pseudo-spinor states, these operators may be defined in different forms and may not be realized as differential operators, as occurs for the non-relativistic problem. These facts would lead to some special features of graphene coherent states that are not observed in analogous one-component non-relativistic systems.

We have constructed two families of partial coherent states $\Pi_{\beta,n}$ and $\Pi_{m,\alpha}$ as eigenstates of each annihilation operator together with a complementary number operator. We have also obtained the two-dimensional coherent states $\Upsilon_{\alpha,\beta}$ for graphene, which are common eigenstates of the operators $\mathbb{A}^-$ and $\mathbb{B}^-$. Only the family of coherent states $\Pi_{\beta,n}$ has analytic expression, and their interpretation as displaced states is fully implemented. The other coherent states, although they share the expected properties, do so in a more ``fuzzy'' way. For example, the interpretation of $\Upsilon_{\alpha,\beta}$ as displaced states due to the parameter $\beta$ and having a shape depending on $\alpha$ is correct, but it is not clear how to find a closed expression showing these properties due to the lack of analytic formulas.

Another special feature of the pseudo-spinor eigenstates $\Psi_{m,n}$ is that, except for the ground states where $n=0$, they have a non-vanishing current density or probability flux, which is inherited by the coherent states. In the case of the coherent states $\Pi_{\beta,n}$, there is a flux of probability only in the angular direction, around the point in which the probability density reaches its maximum. The origin of this fact is that the operators $B^{\pm}$ operators implement at the quantum level the integrals of motion that in a classical approach determine the location of the center of the orbit in which a charged particle moves. On the other hand, for the states $\Pi_{m,\alpha}$ and $\Upsilon_{\alpha,\beta}$ there is a flux of probability in the angular and radial directions, without a clear axial symmetry. We assume that this is due to the fact that both quantum states do not have a definite angular momentum and there is no well-defined point where they move.

Although the eigenstates $\Psi_{m,n}$ have components $\psi^u=\psi_{m,n-1}$ and $\psi^d=\psi_{m,n}$ with different orbital angular momentum, $l-1$ and $l$, respectively, the pseudo-spinor is characterized by a well-defined total angular momentum $\mathbb{J}_{z}$ given by $j=l-1/2$. We have achieved the construction of coherent states $\Xi_{j,\zeta}$ with a definite angular momentum in $z$ direction by means of annihilation and creation operators $\mathbb{K}^\pm$ that commute with 
$\mathbb{J}_{z}$ and generate the {\it su}(1,1) algebra \cite{f04,ng03,dhm12,dm13}. We have considered two kinds of coherent states according to the sign of $j$ and for both the probability density has an axial symmetry with respect to the origin of the coordinates. As the values of $\vert j\vert$ increase, the maximum probability amplitude moves radially away from the origin and the same happens with the probability flow in the angular direction (see Figs.~\ref{fig:rho1P_z} and \ref{fig:rho1N_z}). As expected, the flux of probability in the radial direction is negligible because, in a classical interpretation, this situation corresponds to a particle confined to moving in a circular path centered at the origin and whose radius increases with increasing angular momentum.

On the other hand, the analysis of the circular motion through the mean values of the matrix operators in Eq.~(\ref{matrixoperators}) for each coherent state considered in this work has allowed us to obtain a physical interpretation of the eigenvalues $\alpha$, $\beta$ and $\zeta$. Regarding the average energy, for any of the coherent states found here, this is a continuous function of the corresponding eigenvalue, which helps us make a semi-classical interpretation of these quantum states. However, for the {\it su}(1,1) coherent states in graphene it is important to remark the behavior of the mean energy as the angular momentum changes: in Figure \ref{fig:H_z} we have seen that the function $\langle H_{\rm DW}\rangle_\zeta$ takes smaller values as the $z$ component of the total angular momentum $j$ increases. It is worth to remark that, although the probability densities of the states $\Pi_{m,\alpha}$, $\Pi_{\beta,n}$, $\Upsilon_{\alpha,\beta}$ and $\Xi_{j,\zeta}$, as well as the corresponding mean energy values, were plotted for a specific magnetic field strength, our findings can be extended to any other value of the magnetic field $B_{0}$, adjusting the graph scales.

Finally, let us mention that the annihilation operators $\mathbb{A}^{-}$, $\mathbb{B}^-$ and $\mathbb{K}^-$ do not have a unique form. As it was shown in \cite{df17}, it is posible to obtain coherent states associated to operators that generate nonlinear algebras \cite{mmsz93,mmsz93a,hh02,rr00,rr00a,s00}. The possibility of constructing other coherent states generalizations for {\it su}(1,1) and {\it su}(2) algebras \cite{diaz20,f04,ng03,dhm12,dm13}, based on the annihilation operators defined in this work, is quite promising.

\acknowledgments

This work has been supported by Junta de Castilla y Le\'on and FEDER projects (VA137G18 and BU229P18) and CONACYT (Mexico), project FORDECYT-PRONACES/61533/2020. EDB also acknowledges the warm hospitality at Department of Theoretical Physics of the University of Valladolid, as well his family moral support, specially of Act. J. Manuel Zapata L.

\appendix

\section{Densities  and currents for the coherent states $\Pi_{m,\alpha}$}\label{Appendix}
The expressions for the probability densities $\rho_{m,\alpha}$ in the coherent states (\ref{78}) are 

\begin{eqnarray*}
 	\nonumber \rho_{m,\alpha}(\xi,\theta)&=&\Pi^{\dagger}_{m,\alpha} \Pi_{m,\alpha} =\frac1{2\exp(\vert\tilde{\alpha}\vert^2)-1}\Bigg[\vert g_m(\xi)\vert^2+\left\vert\sum_{n=m+1}^{\infty}\frac{\left(\tilde{\alpha}z\right)^n}{n!}L_{m}^{n-m}(\xi^2)f_m(\xi)\right\vert^2\\
 	\nonumber&&+\left\vert\sum_{n=m+1}^{\infty}\frac{\left(\tilde{\alpha} z\right)^n}{n!}\frac{\sqrt{n}}{z}\, L_{m}^{n-m-1}(\xi^2)f_m(\xi)\right\vert^2+2\,\textrm{Re}\left[\sum_{n=m+1}^{\infty}\frac{(\tilde{\alpha}z)^n}{n!}f_m(\xi)g_m^\ast(\xi)\, L_{m}^{n-m}(\xi^2)\right]\\
 	\nonumber&&+(1-\delta_{0m})\left(\left\vert\sum_{n=1}^{m}\left(-\frac{\tilde{\alpha} }{z^\ast}\right)^nL_{n}^{m-n}(\xi^2)g_m(\xi)\right\vert^2+\left\vert\sum_{n=1}^{m}\left(-\frac{\tilde{\alpha}}{z^\ast}\right)^n\frac{z^\ast}{\sqrt{n}}\,  L_{n-1}^{m-n+1}(\xi^2)g_m(\xi)\right\vert^2\right. \\
	\nonumber&&+2\, \textrm{Re}\left[\sum_{n=1}^{m}\left(-\frac{\tilde{\alpha}}{z^\ast}\right)^nL_n^{m-n}(\xi^2)\vert g_m(\xi)\vert^2+\left(\sum_{n'=1}^{m}\left(-\frac{\tilde{\alpha}^\ast}{z}\right)^{n'}L_{n'}^{m-n'}(\xi^2)g_m^\ast(\xi)\right)\times\right. \\
	\nonumber&&\times\left(\sum_{n=m+1}^{\infty}\frac{\left(\tilde{\alpha}z\right)^n}{n!}\, L_{m}^{n-m}(\xi^2)f_m(\xi)\right)-\left(\sum_{n'=1}^{m}\left(-\frac{\tilde{\alpha}^\ast}{z}\right)^{n'}\frac{1}{\sqrt{n'}}\, L_{n'-1}^{m-n'+1}(\xi^2)g_m^\ast(\xi)\right)\times \\
	&&\times\left.\left.\left(\sum_{n=m+1}^{\infty}\frac{\left(\tilde{\alpha}z\right)^n}{n!}\sqrt{n}\, L_{m}^{n-m-1}(\xi^2)f_m(\xi)\right)\right]\right)\Bigg].\label{80a} 
	\\ [2ex]
\end{eqnarray*}
For the current densities $j_{m,\alpha,\vec{u}}$ in the coherent states (\ref{78}) we get
\begin{eqnarray*}
	\nonumber  j_{m,\alpha,\vec{u}}(\xi)&=&ev_{\rm F}\ \Pi^{\dagger}_{m,\alpha} \,(\vec{\sigma}\cdot\vec{u})_k\,
	\Pi_{m,\alpha} \\
	\nonumber&=&\frac{2ev_{\rm F}}{2\exp(\vert\alpha\vert^2)-1}\, \textrm{Re}\Bigg[i(-i)^{k}e^{-i\theta}\left\{\sum_{n=m+1}^{\infty}\frac{(\tilde{\alpha}^\ast z^\ast)^{n}}{n!}\frac{\sqrt{n}}{z^\ast}\, L_{m}^{n-m-1}(\xi^2)f_m^\ast(\xi)g_m(\xi)\right.\\
	\nonumber&&+\left(\sum_{n'=m+1}^{\infty}\frac{(\tilde{\alpha}z)^{n'}}{n'!}\, L_{m}^{n'-m}(\xi^2)f_m(\xi)\right)\left(\sum_{n=m+1}^{\infty}\frac{(\tilde{\alpha}^\ast z^\ast)^{n}}{n!}\frac{\sqrt{n}}{z^\ast}\,  L_{m}^{n-m-1}(\xi^2)f_m(\xi)\right)\\
	\nonumber&&-(1-\delta_{0m})\left(\sum_{n=1}^{m}\left(-\frac{\tilde{\alpha}^\ast}{z}\right)^n\frac{z}{\sqrt{n}}\, L_{n-1}^{m-n+1}(\xi^2)\vert g_m(\xi)\vert^2\right. \\
	\nonumber&&+\left(\sum_{n'=1}^{m}\left(-\frac{\tilde{\alpha}}{z^\ast}\right)^{n'}L_{n'}^{m-n'}(\xi^2)g_m(\xi)\right)\left(\sum_{n=1}^{m}\left(-\frac{\tilde{\alpha}^\ast}{z}\right)^n\frac{z}{\sqrt{n}}\, L_{n-1}^{m-n+1}(\xi^2)g_m^\ast(\xi)\right) \\
	\nonumber&&-\left(\sum_{n'=1}^{m}\left(-\frac{\tilde{\alpha}}{z^\ast}\right)^{n'} L_{n'}^{m-n'}(\xi^2)g_m(\xi)\right)\left(\sum_{n=m+1}^{\infty}\frac{(\tilde{\alpha}^\ast z^\ast)^{n}}{n!}\frac{\sqrt{n}}{z^\ast}\, L_{m}^{n-m-1}(\xi^2)f_m^\ast(\xi)\right) \\
	&&+\left.\left.\left(\sum_{n'=m+1}^{\infty}\frac{(\tilde{\alpha}z)^{n'}}{n'!}\, L_{m}^{n'-m}(\xi^2)f_m(\xi)\right)\left(\sum_{n=1}^{m}\left(-\frac{\tilde{\alpha}^\ast}{z}\right)^{n}\frac{z}{\sqrt{n}}\, L_{n-1}^{m-n+1}(\xi^2)g_m^\ast(\xi)\right)\right)\right\}\Bigg]. \label{80b}
\end{eqnarray*}
In all the cases $z$ is the complex parameter defined in Eq.~(\ref{59}) and
\begin{equation*}
	f_{m}(\xi)=\sqrt{\frac{m!}{2\pi\ell_{\rm B}^{2}}}\,(-z)^{-m}\exp\left(-\frac{1}{2}\xi^2\right), \qquad 
	g_{m}(\xi)=\sqrt{\frac{1}{2\pi\ell_{\rm B}^{2}\,m!}}\,z^{\ast m}\exp\left(-\frac{1}{2}\xi^2\right).
\end{equation*}
Some plots of the functions $\rho_{m,\alpha}$ and $j_{m,\alpha,\vec{u}}$ can be seen on Figure \ref{fig:rho_alpha1}.


\begin{thebibliography}{10}

\bibitem{h93}
P.~Hawrylak, Phys. Rev. Lett. {\bf 71}, 3347 (1993)

\bibitem{kat01}
L.P. Kouwenhoven, D.G. Austing,  S. Tarucha, Rep. Prog.  Phys. {\bf 64}, 701 (2001)

\bibitem{mc94}
A.V. Madhav, T. Chakraborty, Phys. Rev. B {\bf 49}, 8163 (1994)

\bibitem{cac07}
H.-Y. Chen, V. Apalkov,  T. Chakraborty, Phys. Rev. Lett. {\bf 98}, 186803 (2007)

\bibitem{spba04}
B. Szafran, F.M. Peeters, S. Bednarek,  J. Adamowski, Phys. Rev. B {\bf 69}, 125344 (2004)

\bibitem{f28}
V. Fock,  Zeitschrift f{\"u}r Physik {\bf 47}, 446 (1928)

\bibitem{d31}
C.G. Darwin,  Math. Proc. Cam. Philos.  Soc. {\bf 27}, 86 (1931)

\bibitem{p30}
L. Page,  Phys. Rev. {\bf 36}, 444 (1930)

\bibitem{landau30}
L. Landau, Zeitschrift f{\"u}r Physik {\bf 64}, 629 (1930)

\bibitem{mm69}
I.A. Malkin, V.I. Man'ko,  Sov. Phys. JETP {\bf 28}, 527 (1969)

\bibitem{g63}
R.J. Glauber, Phys. Rev. {\bf 131}, 2766 (1963)

\bibitem{fk70}
A. Feldman,  A.H. Kahn,  Phys. Rev. B {\bf 1}, 4584 (1970)

\bibitem{lms89}
G. Loyola, M. Moshinsky,  A. Szczepaniak, Am. J. Phys. {\bf 57}, 811 (1989)

\bibitem{krp96}
K. Kowalski, J. Rembielinski, L.C. Papaloucas, J. Phys. A Math. Gen. {\bf 29}, 4149 (1996)

\bibitem{sm03}
D. Schuch, M. Moshinsky, J. Phys. A Math. Gen. {\bf 36}, 6571 (2003)

\bibitem{kr05}
K. Kowalski, J. Rembieli{\'{n}}ski, J. Phys. A Math. Gen. {\bf 38}, 8247 (2005)

\bibitem{re08}
M.N. Rhimi, R. El-Bahi, Int. J. Theor. Phys. {\bf 47}, 1095 (2008)

\bibitem{d17}
V.V. Dodonov, in {\em Coherent States and Their Applications: A Contemporary Panorama}, Springer Proceedings in Physics (205), pages 311--338, ed. by  J.-P. Antoine, F. Bagarello, J.-P. Gazeau (Springer, Cham, 2018)

\bibitem{z64}
J. Zak,  Phys. Rev. {\bf 134A}, 1602 (1964)

\bibitem{b64}
E. Brown,  Phys. Rev. {\bf 133A}, 1038 (1964)

\bibitem{l83}
R.B. Laughlin,  Phys. Rev. B {\bf 27}, 3383 (1983)

\bibitem{wz94}
P.B. Wiegmann, A.V. Zabrodin, Phys. Rev. Lett. {\bf 72}, 1890 (1994)

\bibitem{fw99}
M.K. Fung, Y.F. Wang, Chin. J. Phys. {\bf 38}, 10 (2000)

\bibitem{ngmzd04}
K.S. Novoselov, A.K. Geim, S.V. Morozov, D. Jiang, Y. Zhang, S.V. Dubonos, I.V.  Grigorieva,  A.A. Firsov,
 Science {\bf 306}, 666 (2004)

\bibitem{ztsk05}
Y. Zhang, Y.W. Tan, L.S. Horst,  P. Kim, Nature {\bf 438}, 201 (2005)

\bibitem{cngpn09}
A.H. Castro~Neto, F. Guinea, N.M.R. Peres, K.S. Novoselov,  A.K. Geim,
 Rev. Mod. Phys. {\bf 81}, 109 (2009)

\bibitem{kng06}
M.I. Katsnelson, K.S. Novoselov,  A.K. Geim, Nat. Phys. {\bf 2}, 620 (2006)

\bibitem{s91}
A.M.J. Schakel, Phys. Rev. D {\bf 43}, 1428 (1991)

\bibitem{k06}
M.I. Katsnelson,  Eur. Phys. J. B {\bf 51}, 157 (2006)

\bibitem{rz07}
T.M. Rusin, W. Zawadzki,  Phys. Rev. B {\bf 76}, 195439 (2007)

\bibitem{rz08}
T.M. Rusin, W. Zawadzki,  Phys. Rev. B {\bf 78}, 125419 (2008)

\bibitem{dmdae07}
A. De Martino, L. Dell'Anna,  R. Egger,  Phys. Rev. Lett. {\bf 98}, 066802 (2007)

\bibitem{dndm09}
L. Dell'Anna, A. De~Martino,  Phys. Rev. B {\bf 79}, 045420 (2009)

\bibitem{gmr09}
G. Giavaras, P.A. Maksym,  M. Roy, J. Phys. Condens. Matter {\bf 21}, 102201 (2009)

\bibitem{knn09}
\c{S}. Kuru, J. Negro, L.M. Nieto, J. Phys. Condens. Matter  {\bf 21}, 455305 (2009)

\bibitem{mf14}
B. Midya, D.J. Fern{\'{a}}ndez, J. Phys. A. Math. Theor. {\bf 47}, 285302 (2014)

\bibitem{rvp11}
M. Ramezani Masir, P. Vasilopoulos,  F.M. Peeters, J. Phys. Condens. Matter {\bf 23}, 315301 (2011)

\bibitem{lara}
M. Gadella, L.P. Lara, J. Negro, Int. J. Mod. Phys. C {\bf 28}, 1750036 (2017)

\bibitem{dp16}
C.A. Downing, M.E. Portnoi,  Phys. Rev. B  {\bf 94}, 165407 (2016)

\bibitem{cdmp16}
C.A. Downing, M.E. Portnoi,  Phys. Rev. B  {\bf 94}, 045430 (2016)

\bibitem{ema17}
M. Eshghi, H. Mehraban,  I.A. Azar,  Physica E  {\bf 94}, 106 (2017)

\bibitem{rkb12}
P. Roy, T. Kanti Ghosh,  K. Bhattacharya, J. Phys. Condens. Matter {\bf 24}, 055301 (2012)

\bibitem{dnvhp17}
D.N. Le, V.-H. Le,  P. Roy, Physica E {\bf 96}, 17 (2018)

\bibitem{df17}
E. D{\'i}az-Bautista, D.J. Fern{\'a}ndez, Eur. Phys. J. Plus {\bf 132}, 499 (2017)

\bibitem{bg71}
A.O. Barut, L. Girardello,  Commun. Math. Phys. {\bf 21}, 41 (1971)

\bibitem{dknn17}
E. Drigho-Filho, \c{S}. Kuru, J. Negro,  L.M. Nieto, Ann. Phys. {\bf 383}, 101 (2017)

\bibitem{df96}
D.J. Fern\'andez C., J. Negro,  M.A. del Olmo, Ann. Phys. {\bf 252}, 386  (1996)

\bibitem{kka12}
K. Kikoin, M. Kiselev,  Y. Avishai, {\em Dynamical Symmetries in Molecular Electronics} (Springer, Vienna, 2012), pp. 197–231

\bibitem{s18}
L. Sourrouille, J. Phys. Commun. {\bf 2}, 045030 (2018)

\bibitem{kns18}
\c{S}. Kuru, J. Negro,  L. Sourrouille, J. Phys. Condens. Matter {\bf 30}, 365502 (2018)

\bibitem{ac79}
Y. Aharonov, A. Casher, Phys. Rev. A {\bf 19}, 2461 (1979)

\bibitem{w81}
E. Witten,  Nucl. Phys. B  {\bf 188}, 513 (1981)

\bibitem{az86}
C. Aragone, F. Zypman, J. Phys. A Math. Gen. {\bf 19}, 2267 (1986)

\bibitem{bh93}
Y. B\'erub\'e-Lauzi\`{e}re, V. Hussin, J. Phys. A Math. Gen. {\bf 26}, 6271 (1993)

\bibitem{kz13}
M. Kornbluth, F. Zypman, J. Math. Phys. {\bf 54}, 012101 (2013)

\bibitem{diaz20}
E. D{\'i}az-Bautista,  J. Math. Phys. {\bf  61}, 102101 (2020)

\bibitem{dosr20}
E. D{\'i}az-Bautista, M. Oliva-Leyva, Y. Concha-S{\'a}nchez, A. Raya, J. Phys. A Math. Theor. {\bf  53}, 105301 (2020)

\bibitem{mmsz93}
V.I. Man'ko, G. Marmo, S. Solimeno,  F. Zaccaria, Int. J. Mod. Phys. A  {\bf 08}, 3577 (1993)

\bibitem{mmsz93a}
V.I. Man'ko, G. Marmo, S. Solimeno,  F. Zaccaria, Phys. Lett. A  {\bf  176}, 173 (1993)

\bibitem{hh02}
F. Hong-Yi, C. Hai-Ling, Commun. Theor. Phys.  {\bf 37}, 655 (2002)

\bibitem{rr00}
B. Roy, P. Roy, J. Opt. B  {\bf 2}, 65 (2000)

\bibitem{rr00a}
B. Roy, P. Roy, J. Opt. B  {\bf 2}, 505 (2000)

\bibitem{s00}
S. Sivakumar, J. Opt. B  {\bf 2}, R61 (2000)

\bibitem{p72}
A.M. Perelomov, Commun. Math. Phys.  {\bf 26}, 222 (1972)

\bibitem{bbdr76}
D. Bhaumik, K. Bhaumik,  B. Dutta-Roy, J. Phys. A Math. Gen.  {\bf 9},  1507 (1976)

\bibitem{f04}
H. Fakhri, J. Phys. A Math. Gen.  {\bf 37},  5203 (2004)

\bibitem{ahb15}
I. Aremua, M.N. Hounkonnou, E. Balo\"{i}tcha, Rep. Math. Phys. {\bf 76}, 247 (2015)


\bibitem{london50}
F. London, {\it Superfluids}, Wiley, New York (1950).

\bibitem{onsager54}
L. Onsager, in: {\it Proceedings of the International Conference on Theoretical Physics} (Kyoto \& Tokyo, September 1953), Science Council of Japan, Tokyo (1954), p. 935.

\bibitem{jacak20} J.E. Jacak, New J. Phys. {\bf 22}, 093027 (2020)

\bibitem{ng03}
M. Novaes,  J.P. Gazeau, J. Phys. A Math. Gen.  {\bf 36}, 199 (2002)

\bibitem{dhm12}
A. Dehghani, H. Fakhri, B. Mojaveri, J. Math. Phys. {\bf 53}, 123527 (2012)

\bibitem{dm13}
A. Dehghani, B. Mojaveri, Eur. Phys. J. D {\bf 67}, 264 (2013)

\end{thebibliography}
\end{document}